\documentclass[a4paper,fleqn,usenatbib]{mnras}

\usepackage{amsmath}
\usepackage{mathptmx}
\usepackage{txfonts}
\usepackage{lscape}
\usepackage{tabularx}
\usepackage{stfloats}

\usepackage[T1]{fontenc}
\usepackage{ae,aecompl}

\usepackage{graphicx}   

\title[BDBS Paper IV]{Blanco DECam Bulge Survey (BDBS) IV: Metallicity
Distributions and Bulge Structure from 2.6 Million Red Clump Stars}
\author[Johnson et al.]{
Christian I. Johnson,$^{1}$\thanks{E-mail: chjohnson1@stsci.edu}
R. Michael Rich,$^{2}$
Iulia T. Simion,$^{3}$
Michael D. Young,$^{4}$
\newauthor
William I. Clarkson,$^{5}$
Catherine A. Pilachowski,$^{6}$
Scott Michael,$^{6}$
\newauthor
Tommaso Marchetti,$^{7}$
Mario Soto,$^{8}$
Andrea Kunder,$^{9}$
Andreas J. Koch-Hansen,$^{10}$
\newauthor
A. Katherina Vivas,$^{11}$
Meridith Joyce,$^{1,15}$
Juntai Shen,$^{12,13}$
and
Alexis Osmond$^{5,14}$
\\
$^{1}$Space Telescope Science Institute, 3700 San Martin Drive, Baltimore, MD 
21218, USA\\
$^{2}$Department of Physics and Astronomy, UCLA, 430 Portola Plaza,
Box 951547, Los Angeles, CA 90095-1547, USA\\
$^{3}$Shanghai Key Lab for Astrophysics, Shanghai Normal University, 100 Guilin Road, Shanghai, 200234\\
$^{4}$Indiana University, University Information Technology Services, CIB 2709 
E 10th Street, Bloomington, IN 47401 USA\\
$^{5}$Department of Natural Sciences, University of Michigan-Dearborn, 4901 Evergreen Rd. Dearborn, MI 48128, USA\\
$^{6}$Indiana University Department of Astronomy, SW319, 727 E 3rd Street, Bloomington, IN 47405 USA\\
$^{7}$European Southern Observatory, Karl-Schwarzschild-Strasse 2, 85748 Garching bei München, Germany\\
$^{8}$Instituto de Astronomía y Ciencias Planetarias, Universidad de Atacama, Copayapu 485, Copiapó, Chile\\
$^{9}$Saint Martin's University, 5000 Abbey Way SE, Lacey, WA 98503, USA\\
$^{10}$Zentrum f\"ur Astronomie der Universit\"at Heidelberg, Astronomisches 
Rechen-Institut, M\"onchhofstr. 12, 69120 Heidelberg, Germany\\
$^{11}$Cerro Tololo Inter-American Observatory, NSF's National Optical-Infrared Astronomy Research Laboratory, Casilla 603, La Serena, Chile\\
$^{12}$Department of Astronomy, School of Physics and Astronomy, Shanghai Jiao Tong University, 800 Dongchuan Road, Shanghai 200240\\
$^{13}$Key Laboratory for Particle Astrophysics and Cosmology (MOE) / Shanghai Key Laboratory for Particle Physics and Cosmology, \\
Shanghai 200240, China\\
$^{14}$Department of Physics and Astronomy, University of South Carolina, 712 Main St., Room 404, Columbia, S.C. 29208, USA \\
$^{15}$Lasker Fellow
}

\date{Accepted XXX. Received YYY; in original form ZZZ}

\pubyear{2021}

\begin{document}
\label{firstpage}
\pagerange{\pageref{firstpage}--\pageref{lastpage}}
\maketitle
\begin{abstract}
We present photometric metallicity measurements for a sample of 2.6 million 
bulge red clump stars extracted from the Blanco DECam Bulge Survey (BDBS).  
Similar to previous studies, we find that the bulge exhibits a strong vertical 
metallicity gradient, and that at least two peaks in the metallicity 
distribution functions appear at $b < -5^{\circ}$.  We can discern a 
metal-poor ([Fe/H] $\sim$ $-0.3$) and metal-rich ([Fe/H] $\sim$ $+$0.2) 
abundance distribution that each show clear systematic trends with latitude, 
and may be best understood by changes in the bulge's star formation/enrichment 
processes.  Both groups exhibit asymmetric tails, and as a result 
we argue that the proximity of a star to either peak in [Fe/H] space is not 
necessarily an affirmation of group membership.  The metal-poor peak shifts to 
lower [Fe/H] values at larger distances from the plane while the metal-rich 
tail truncates.  Close to the plane, the metal-rich tail appears broader along 
the minor axis than in off-axis fields.  We also posit that the bulge has two 
metal-poor populations -- one that belongs to the metal-poor tail of the low 
latitude and predominantly metal-rich group, and another belonging to the 
metal-poor group that dominates in the outer bulge.  We detect the X-shape 
structure in fields with |Z| $>$ 0.7 kpc and for stars with [Fe/H]~$>$~$-0.5$.  
Stars with [Fe/H] $<$ $-0.5$ may form a spheroidal or ``thick bar" distribution 
while those with [Fe/H] $\ga$ $-$0.1 are strongly concentrated near the plane. 

\end{abstract}




\section{Introduction} \label{sec:intro}

Bulge formation in massive disk galaxies is a complicated process that can
involve a variety of physical mechanisms, such as: violent dissipative collapse
\citep{Eggen62,Larson74}, merger events \citep[][]{Cole00,Hopkins09,
Athanassoula17}, secular disk evolution \citep{Combes90,Kormendy04}, clump
coalescence \citep{Elmegreen08}, and early gas-compaction \citep{Dekel14,
Zolotov15,Tacchella16}.  These processes, or some combination thereof, 
result in two broad classes of bulges: classical and pseudobulges 
\citep{Kormendy04,Athanassoula05,Fisher16}.  However, some galaxies are known
to harbor both components simultaneously \citep{Peletier07,Erwin15,Erwin21}.

Classical/merger-built bulges form violently and early in a galaxy's history,
and produce spheroidal, pressure-supported structures dominated by old stars
that can also exhibit a metallicity gradient.  Conversely, pseudobulges form
over longer time scales, exhibit cylindrical rotation, have flatter and 
more elongated profiles, may contain a broad age dispersion, and can display a 
boxy/peanut X-shape structure when viewed at appropriate angles.  Although the
Milky Way bulge exhibits some classical bulge characteristics, such as the 
existence of a vertical metallicity gradient \citep[e.g.,][]{Zoccali08,
Johnson11,Gonzalez11,Gonzalez13,Zoccali17,Rojas20} and a predominantly old
age \citep[e.g.,][]{Ortolani95,Zoccali03,Clarkson08,Valenti13,Renzini18,
Surot19,Sit20}, the clear signatures of cylindrical rotation \citep{Howard09,
Kunder12,Ness13_kinematic,Zoccali14} and an extended X-shape structure 
\citep{McWilliam10,Nataf10,Saito11,Gonzalez15_X,Ness16} are indicative of 
a dominant pseudobulge/bar.

Interestingly, the Milky Way bulge also exhibits strong evidence supporting a 
composite system.  For example, stars with [Fe/H] $<$ $-$0.5 do not follow the 
same bar-like cylindrical rotation pattern observed in more metal-rich stars, 
and instead exhibit slow or null rotation and high velocity 
dispersion that are more reminiscent of a kinematically hot bar or spheroidal 
population \citep{Soto07,Babusiaux10,Dekany13,Ness13_kinematic,Rojas14,Kunder16,
Rojas17,Zoccali17,Clarkson18,Arentsen20,Kunder20,Rojas20,Wylie21}.  Additional 
evidence supporting an accreted, rather than completely \emph{in situ}, 
population is indicated by the existence of: at least two distinct RR Lyrae 
populations with different kinematic and pulsation period properties 
\citep{Soszynski14,Pietrukowicz15,Kunder19,Kunder20}, N-rich \citep{Schiavon17,
Horta21,Kisku21} and Na-rich \citep{Lee19} stars that are common in globular 
clusters but not in the field, and peculiar clusters such as Terzan 5 
\citep{Ferraro09,Ferraro16} and Liller 1 \citep{Ferraro21} that have large 
metallicity and/or age spreads.  The latter clusters are posited to be remnant
building blocks of the inner Galaxy; however, kinematic simulations suggest 
that any contribution to the total bulge mass by accretion/merger processes
should be $<$ 8-10 per cent of the disk mass \citep{Shen10}.

The composite nature of the Milky Way bulge/bar system is further confused by
discrepancies in age measurements.  Color-magnitude diagram (CMD) analyses 
almost universally find that bulge stars are $>$10 Gyr in age with a 
relatively small ($\sim$1-2 Gyr) age spread \citep{Ortolani95,Kuijken02,
Zoccali03,Clarkson08,Valenti13,Gennaro15,Renzini18}, and its RR Lyrae 
population may be among the oldest in the Galaxy \citep{Savino20}.  In 
contrast, several spectroscopic analyses \citep[e.g.,][]{Bensby13,Bensby17,
Bovy19,Hasselquist20} have found that the bulge, particularly close to the 
plane, hosts a significant fraction of stars with ages $\sim$2-8 Gyr.  
\citet{Haywood16} further claim that a degeneracy between age and metallicity 
could be masking the presence of young stars in previous CMD analyses, and 
that a substantial young population may be required to explain the relatively 
narrow main-sequence turn-off color spreads.  However, the multi-color analysis
by \citet{Renzini18} argues against a significant ($>$ 5 per cent) 
population of young ($<$ 5 Gyr) stars in the bulge, even in relatively low
latitude fields.  Additionally, \citet{Barbuy18} noted 
that the CMD simulation in \citet{Haywood16} that included a prominent young 
population produced too many bright main-sequence turn-off stars and thus may 
be incompatible with observations.  A similar discrepancy is seen in the 
possible discovery of young ($\sim$1 Gyr) blue loop stars in Baade's Window by 
\citet{Saha19}, which are shown by \citet{Rich20} to instead be nearby stars 
in the disk.

Uncertainty also surrounds the different metallicity distribution function 
interpretations and links between the bulge and thin/thick disk stars based on
detailed chemical composition comparisons.  Numerous sources find multiple 
``peaks" in the metallicity distribution functions that vary in amplitude as a 
function of Galactic latitude \citep[e.g.,][]{Hill11,Bensby13,Ness13_mdf,
Bensby17,Zoccali17,Duong19,Johnson20,Rojas20,Wylie21}, but both the number of 
populations fit and the [Fe/H] centroid locations of each group are 
inconsistent between papers.  The discrepant results, coupled with 
significant differences in measurement methods and metallicity scales, 
complicate efforts to map the identified groups into known (e.g., 
thin/thick disk; halo) or newly defined (e.g., old classical bulge) 
populations.

Many papers speculate that the ``metal-poor" and ``metal-rich" peaks represent 
an inner extension of the thick and thin disks into the bulge/bar region 
\citep[e.g.,][]{Ness13_mdf,DiMatteo14,DiMatteo15,Debattista17,Fragkoudi18,
DiMatteo19}.  Such an assertion is supported by the similar detailed abundance 
patterns found in bulge stars and local thin/thick disk stars 
\citep[e.g.,][]{Melendez08,AlvesBrito10,Gonzalez11,Jonsson17,Zasowski19}.  
However, other surveys frequently find chemical evidence that especially the 
metal-poor bulge stars were enriched more quickly than the local disk 
\citep[e.g.,][]{Zoccali06,Fulbright07,Johnson11,Bensby13,Johnson14,Bensby17,
Rojas17,Duong19_chem}.

Several of the issues outlined above result from the bulge's complex 
geometric structure, strong differential reddening, high stellar density, 
significant foreground contamination, and large projection on the sky
\citep[e.g.,][]{Gonzalez12,Gonzalez18}.  A variety of large-scale surveys, 
such as the Bulge Radial Velocity Assay \citep[BRAVA;][]{Rich07,Kunder12}, 
Vista Variables in the Via Lactea \citep[VVV;][]{Minniti10}, Gaia-ESO Survey 
\citep{Gilmore12}, Abundances and Radial velocity Galactic Origins Survey 
\citep[ARGOS;][]{Freeman13}, GIRAFFE Inner Bulge Survey 
\citep[GIBS;][]{Zoccali14}, Apache Point Observatory Galactic Evolution 
Experiment \citep[APOGEE;][]{Majewski17}, HERMES Bulge Survey
\citep[HERBS;][]{Duong19}, and A2A survey \citep{Wylie21}, have made 
significant progress in obtaining chemodynamic data in the bulge.  However,
the aforementioned problems have generally limited spectroscopic and/or 
photometric [Fe/H] measurements to $\sim$10$^{\rm 4}$ total stars and often only
$\sim$100-200 stars per field.  

A new approach utilizing the $u-i$ color and the 3 square degree field-of-view
of the Dark Energy Camera \citep[DECam;][]{Flaugher15} was demonstrated by
\citet{Johnson20} to be an efficient method for measuring [Fe/H] in large
samples of bulge red clump stars.  The color-metallicity relation presented in 
\citet{Johnson20} provides the framework for expanding the number of red 
clump stars with [Fe/H] measurements from thousands to millions.  Therefore,
in this paper we exploit the $\sim$250 million star $ugrizY$ catalog spanning
$>$ 200 square degrees in the Southern Galactic bulge provided by the Blanco
DECam Bulge Survey \citep[BDBS;][]{Rich20,Johnson20} to extract metallicity
and distance estimates for $\sim$2.6 million red clump stars between 
$|l| < 10^{\circ}$ and $-10^{\circ} < b < -3.5^{\circ}$.

\section{BDBS Data Selection} \label{sec:data_selection}
\begin{figure*}
\includegraphics[width=\textwidth]{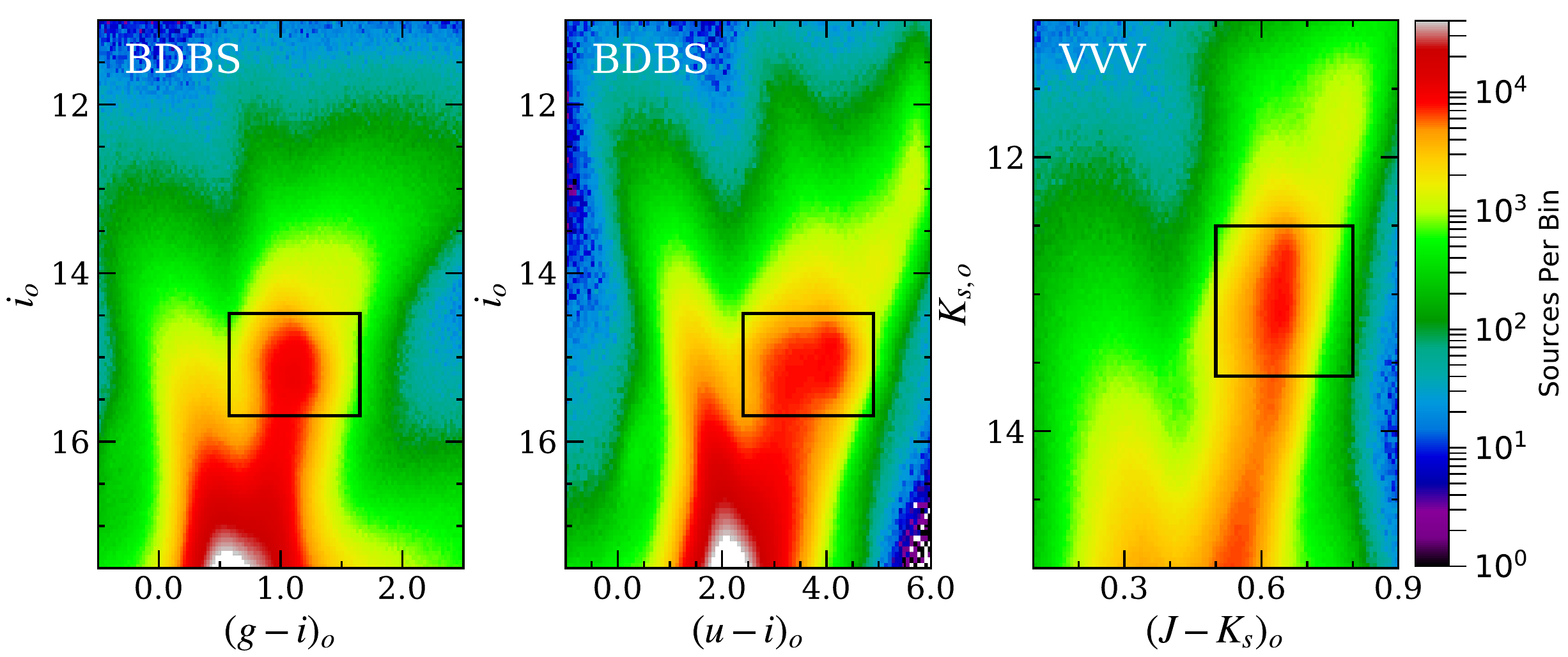}
\caption{Density maps for dereddened CMDs from BDBS (left and middle panels) 
and VVV (right panel) are shown for fields with $|l| < 10^{\circ}$ and 
$-10^{\circ} < b < -3.5^{\circ}$.  The black boxes indicate the adopted 
red clump regions in each CMD.  The blue plumes to the left of the black boxes 
are predominantly composed of foreground disk stars.  Note that while the 
$i_{o}$ vs. $(g-i)_{o}$ and $K_{s,o}$ vs. $(J-K_{s})_{o}$ CMDs exhibit 
relatively narrow red clump color dispersions, the $i_{o}$ vs. $(u-i)_{o}$ red 
clump spans at least 2 mag. in color.  The $(u-i)_{o}$ red clump spread is 
driven by metallicity variations, and highlights the utility of the $u$-band
over redder filters in separating stars by heavy element composition.}
\label{fig:rc_cmd}
\end{figure*}

Low mass stars with [Fe/H] $\ga$ $-1$ evolve off the red giant branch (RGB) 
after undergoing core He ignition and settle into the stable red clump 
evolutionary phase.  Red clump stars are easily observed in the bulge (see 
Figure \ref{fig:rc_cmd}) and serve as standard candles for tracing the three 
dimensional structure of the inner Galaxy \citep[e.g.,][]{Stanek94,McWilliam10,
Nataf10,Saito11,Cao13,Wegg13,Simion17,Gonzalez18,Paterson20}.  Although the red 
clump feature stands out as a clear over-density on the blue side of the RGB in 
Figure \ref{fig:rc_cmd}, isolating a pure bulge red clump sample can be 
challenging.  This is particularly true for the BDBS catalog adopted
here, which uses near-UV and optical filters, covers a large area, and includes
hundreds of millions of stars.

Several effects conspire to add uncertainty in the red clump selection process,
such as: differential reddening, distance variations, and foreground 
contamination along a line-of-sight.  Differential reddening was
addressed by employing the 1$\arcmin$ $\times$ 1$\arcmin$ reddening map 
described in \citet{Simion17} and \citet{Johnson20} to correct the observed 
$ugi$ photometry for all fields with $|l| < 10^{\circ}$ and $-10^{\circ} < b < 
-3.5^{\circ}$.  The BDBS footprint extends outside this selection box, but 
our reddening map currently only covers $l$ $\pm$ $10^{\circ}$ and $b > 
-10^{\circ}$.  Fields interior to $b = -3.5^{\circ}$ are also omitted due to 
significantly higher extinction and stronger sub-arcminute reddening 
variations along most lines-of-sight.

For the dereddened $ugi$ BDBS CMDs utilized in this work, distance variations 
among red clump stars are manifested as a vertical spread in magnitude.  
Accounting for distance variations during the target selection process is 
particularly important because: (1) the bulge/bar system is oriented at an
angle of $\sim$ 30$^{\circ}$ and thus stars are distributed at distances 
exceeding $\pm$ 1 kpc from the Galactic Center, depending on the sight line; 
and (2) the bulge's X-shape structure produces a double red clump along some 
sight lines, particularly those on the near-side of the bar with 
$b < -5^{\circ}$ \citep{McWilliam10,Nataf10,Saito11}.

\begin{figure}
\includegraphics[width=\columnwidth]{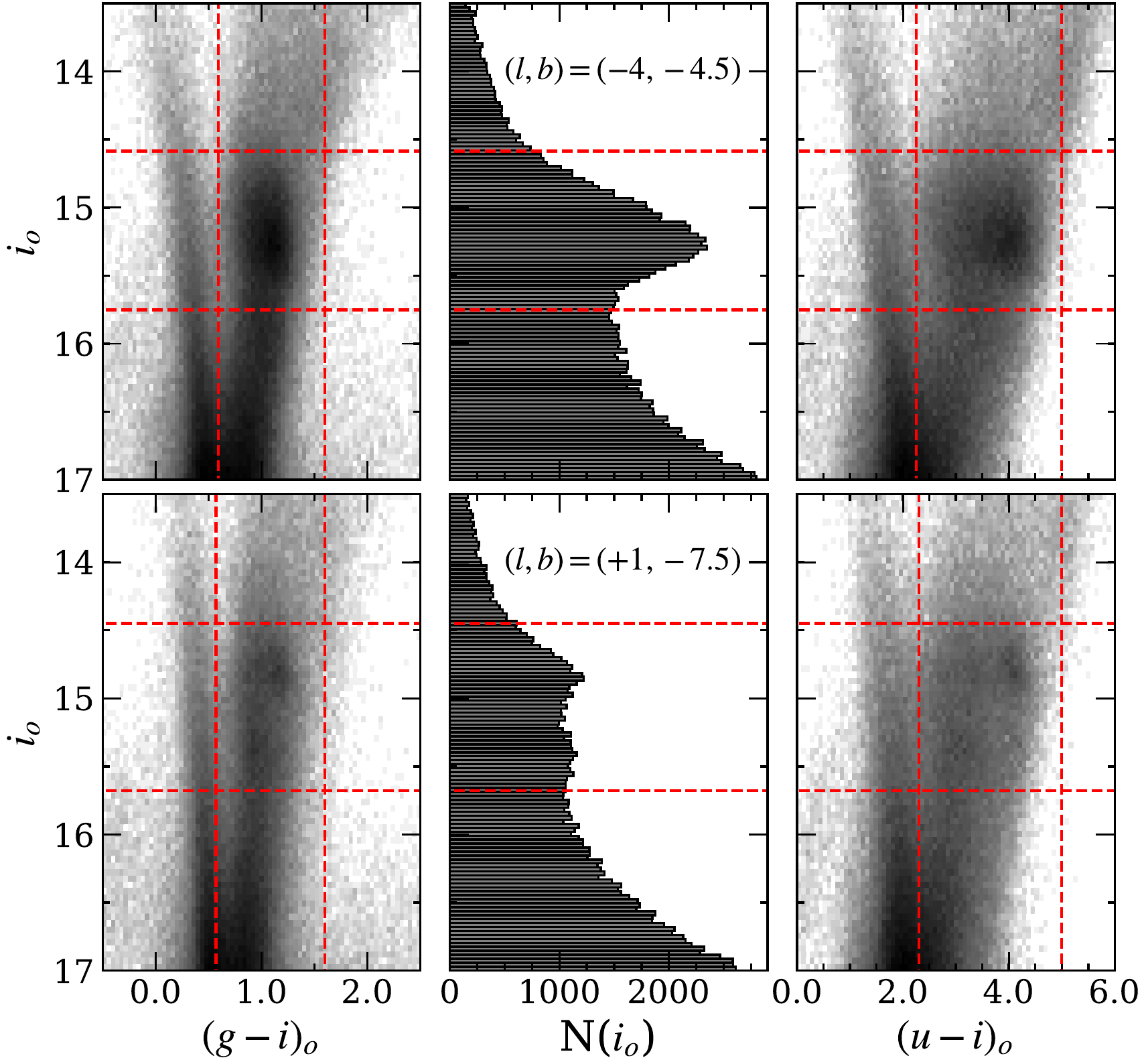}
\caption{The left panels show $i_{o}$ vs. $(g-i)_{o}$ BDBS CMDs for 
0.5$^{\circ}$ $\times$ 0.5$^{\circ}$ fields centered near $(l,b)$ = $(-4,-4.5)$
on the top and $(l,b)$ = $(+1,-7.5)$ on the bottom.  The bulge red clump and 
foreground blue plume populations are obvious in both panels.  The vertical 
and horizontal dashed red lines illustrate the magnitude and color limits for 
the red clump selection boxes in each field.  The middle panels show $i_{o}$ 
luminosity functions for stars inside the selection boxes.  The right panels 
show similar $i_{o}$ vs. $(u-i)_{o}$ CMDs with the dashed red lines indicating 
equivalent selection boxes.  Note the increased color dispersion and complex
morphology present when using the $u$-band, and the double red clump in the
outer bulge field.}
\label{fig:cmd_selection}
\end{figure}

Foreground contamination is mostly observed as an overlap between the nearly
vertical blue plume feature seen in Figure \ref{fig:rc_cmd} and the blue edge 
of the bulge red clump.  Although the bulge red clump and foreground blue plume
features are well-separated along some sight lines, in other fields the 
distinction is more nebulous (see Figure \ref{fig:cmd_selection}).  The 
variable separations between the red clump and foreground disk are largely
driven by changes in the mean metallicity of bulge stars with Galactic latitude
that are coupled with differences between foreground/bulge reddening along a 
particular line-of-sight (see also Section \ref{sec:appendix}).

Given the large amount of data in the BDBS catalog, along with distance and 
contamination effects that need to be controlled, we determined the red clump
selection boxes by visually inspecting $i_{o}$ vs. $(g-i)_{o}$ CMDs in 
3 square degree regions across the aforementioned survey footprint (i.e., the
sky grid boxes shown in Figure 5 of \citealt{Johnson20}).  For each field, the 
bright and faint $i_{o}$ limits were determined by inspecting the $i$-band 
luminosity function and designating where the star counts rose above the 
background RGB distribution.  In fields where the double red clump was 
prominent, the bright and faint limits were set to encapsulate both populations.

Color cuts in each visual inspection field were determined purely from 
$(g-i)_{o}$.  A constant red limit of $(g-i)_{o}$ = 1.6 was used for all 
fields.  However, the blue color limit was set such that the visible portion
of the blue plume did not overlap with the red clump at the faint $i_{o}$ limit.

\begin{figure}
\includegraphics[width=\columnwidth]{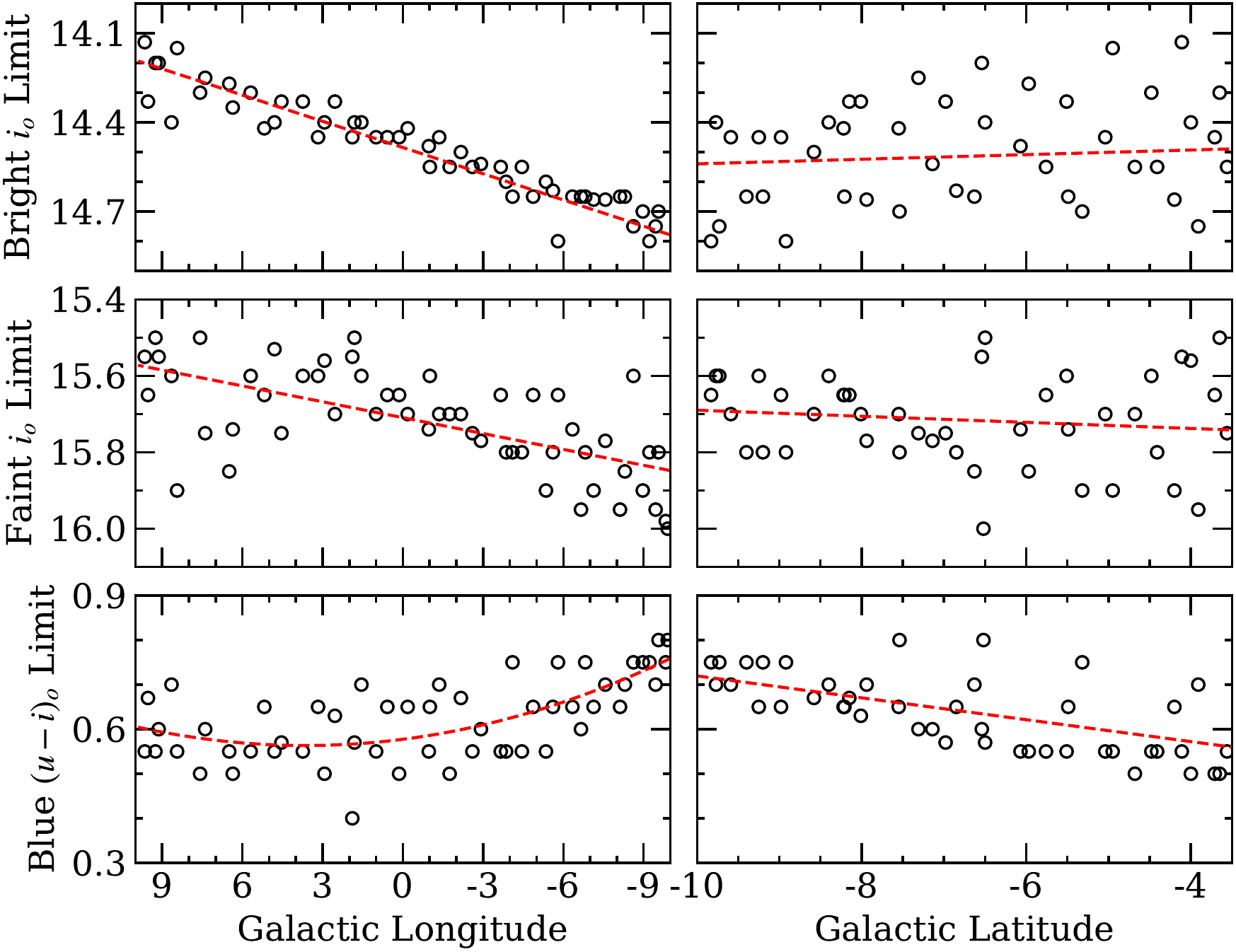}
\caption{The top, middle, and bottom rows show the selection box limits for 
the bright $i_{o}$, faint $i_{o}$, and blue $(u-i)_{o}$ parameters, 
respectively, as measured for individual visually inspected fields (open
circles).  The left column shows the change in these parameters as a function
of Galactic longitude while the right column shows the distribution with
respect to Galactic latitude.  The dashed red lines are low order polynomial
fits to the data.}
\label{fig:limits}
\end{figure}

Since the 3 square degree selection fields are discrete and arbitrarily 
selected while the data are continuous, we fit low order polynomials
to the selection box limits in order to obtain analytical cut-offs as functions
of Galactic latitude and/or longitude.  The data and fits 
are shown in Figure \ref{fig:limits}, and indicate that the bright, faint, 
and blue limits vary smoothly as a function of Galactic longitude.  The bright 
and faint $i$-band limits showed significantly more scatter and lower 
sensitivity to Galactic latitude; however, the blue $(u-i)_{o}$ limit was 
somewhat sensitive to both Galactic latitude and longitude.  For consistency, 
we calculated the selection box limits using the longitude-based ($l$) 
functional forms only, which are:

\begin{equation} \label{eq:bright_lim}
i_{o,bright}=-0.027l+14.477,
\end{equation}

\begin{equation} \label{eq:faint_lim}
i_{o,faint}=-0.015l+15.692, 
\end{equation}
\noindent
and

\begin{equation} \label{eq:blue_lim}
(u-i)_{o,blue}=0.0011l^{2}-0.0077l+0.577,
\end{equation}

with typical uncertainties of about 0.05-0.10 magnitudes.  A star's initial 
selection as a potential bulge red clump member was therefore based solely on 
its Galactic longitude and the application of Equations 
\ref{eq:bright_lim}-\ref{eq:blue_lim} (along with the additional constraint of
$(g-i)_{o} < 1.6$).  Additional examples of the color and magnitude limits 
for various fields, based on applying 
Equations \ref{eq:bright_lim}-\ref{eq:blue_lim}, are described in 
Section \ref{sec:appendix} and 
Figures \ref{fig:app_cmd_selection_grid}-\ref{fig:app_cmd_selection_grid2}.

\begin{figure}
\includegraphics[width=\columnwidth]{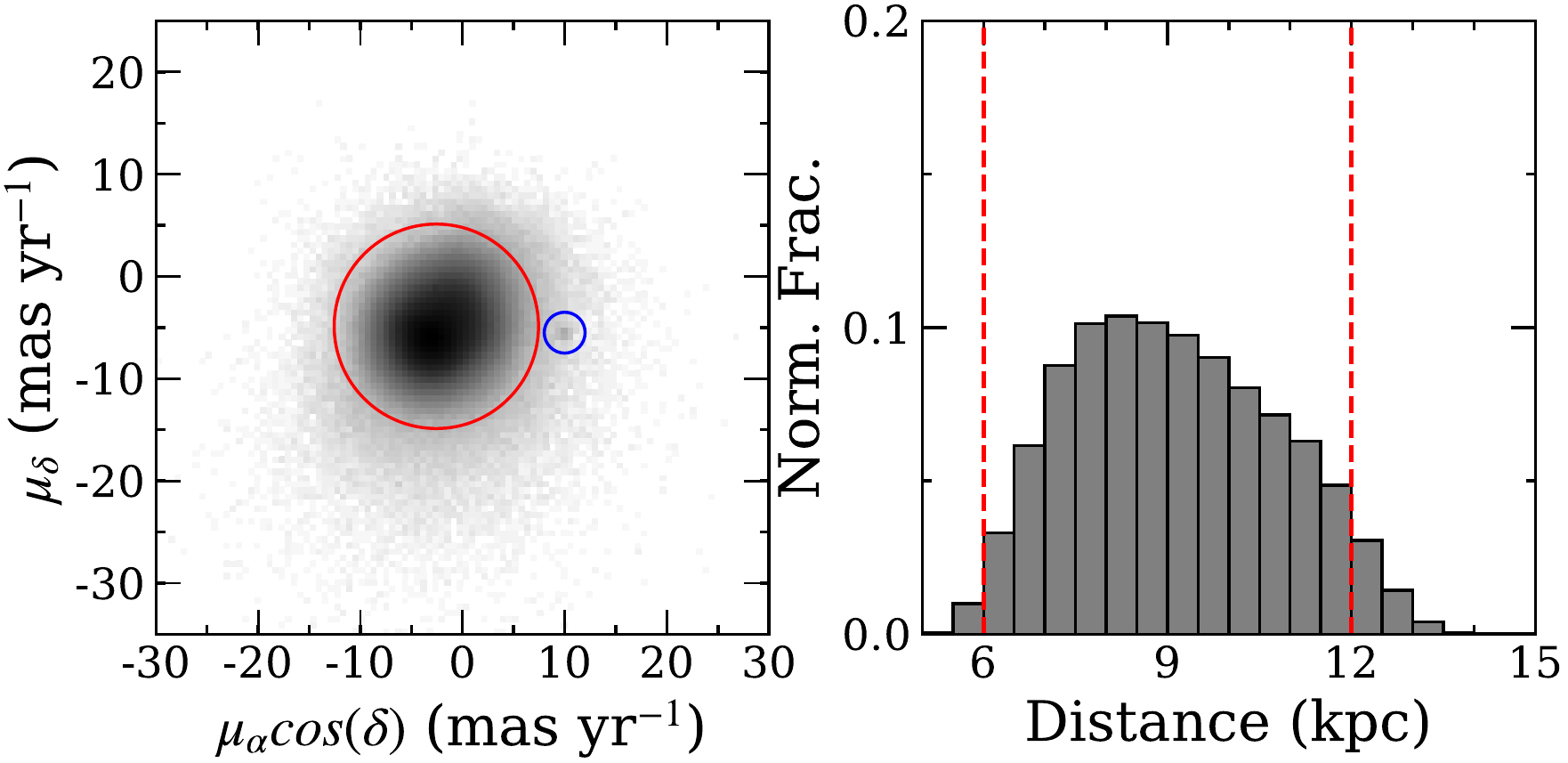}
\caption{\emph{Left:} a vector point proper motion diagram is shown for all 
potential red clump stars in our BDBS sample when matched to Gaia EDR3.  The 
large red circle indicates the selection criteria used to separate likely 
bulge members from foreground field stars.  The red circle has a radius of 9 
mas yr$^{-1}$ ($\sim$ 1$\sigma$) and is centered at $\mu_{\alpha}$cos($\delta$)
= $-$2.562 mas yr$^{-1}$ and $\mu_{\delta}$ = $-$5.376 mas yr$^{-1}$.  The 
small clump observed to the right of the bulge population (blue circle) is the 
residual population of stars in the globular cluster M~22.  \emph{Right:} a 
histogram of red clump distances is shown for stars inside the red circle of 
the left panel.  The peak of the distribution is approximately 8.2 kpc.}
\label{fig:pm_plot}
\end{figure}

The number of stars initially identified as possible bulge red clump members
exceeded 3 million.  However, we culled this list further by removing targets
with low quality measurements (large photometric errors, very high background 
values, unusual shape parameters, etc.) and those with radial distances within
5$\arcmin$ of all known globular clusters.  The 5$\arcmin$ limit is sufficient
to remove stars within $>$2-3 half-light radii of all clusters in our sample
except M~22, which subtends a larger area on the sky.  However, 
Figure \ref{fig:pm_plot} shows that M~22's proper motion is outside our 
selection criteria for bulge field stars, and as a result we do not expect
any significant contamination.  Furthermore, we removed obvious
foreground contamination using proper motions and parallaxes provided by the 
Gaia Early Data Release 3 catalog \citep[EDR3;][]{GaiaEDR3}.  
Figure \ref{fig:pm_plot} shows the results of matching our initial red clump 
catalog against EDR3\footnote{\citet{Rich20} found that Gaia DR2 had a 
completeness limit of $i$ $\sim$ 18 mag. when compared to BDBS in Baade's 
window, which is fainter than the red clump in all fields discussed here.}.  
We only accepted stars inside the large red circle of Figure \ref{fig:pm_plot} 
that also had parallax values below 0.3 mas as bulge members.  

The final step in cleaning the BDBS red clump sample utilized distance cuts
to further constrain likely bulge members.  Although red clump stars are 
reasonably stable standard candles, their $i$-band magnitudes are somewhat 
metallicity dependent.  Numerous red clump absolute magnitude calibrations 
exist for a variety of bands \citep[e.g.,][]{Alves00,Girardi01,Laney12,
Hawkins17,Ruiz18,Onozato19,Plevne20}, but these studies either do not include 
the metallicity dependence or are lacking a calibration for the BDBS bands.  
Therefore, we developed a metallicity dependent $i$-band calibration for stars 
with $-1 <$ [Fe/H] $< +0.5$, based on Pan-STARRS system \citep{Chambers_2016}
isochrones from the MESA Isochrones and Stellar Tracks database
\citep[MIST;][]{Choi16}\footnote{As described in \citet{Johnson20},
the BDBS $u$-band data are calibrated onto the SDSS \citep{Alam_2015} system
while the $grizY$-bands are calibrated onto the Pan-STARRS system.}.  

Briefly, we generated a set of 10 Gyr isochrones with 0.1 dex steps in [M/H]
while assuming [$\alpha$/Fe] = $+$0.3 for [Fe/H] $<$ $-$0.3, 
[$\alpha$/Fe] = $-$[Fe/H] for $-0.3 <$ [Fe/H] $< 0.0$, and [$\alpha$/Fe] = 0.0 
for [Fe/H] $>$ 0.0 \citep[see also][]{Joyce22}.  For compatibility with the
MIST database, we converted the [Fe/H] and [$\alpha$/Fe] values into [M/H] 
ratios via Equation 5 of \citet{Nataf13}:

\begin{equation} \label{eq:mh_conversion}
[M/H] = [Fe/H] + log(0.638 \cdot 10^{[\alpha/Fe]} + 0.362).
\end{equation}
\noindent
We then fit a linear function through the red clump evolutionary point on each 
track where stars spend the most time and derived a relation between absolute 
dereddened $i$-band magnitude ($M_{i,o}$) and [M/H]:

\begin{equation} \label{eq:abs_mag}
M_{i,o} = 0.266[M/H] + 0.364,
\end{equation}
\noindent
or equivalently in [Fe/H] space with the adopted [$\alpha$/Fe] trends folded in:

\begin{equation} \label{eq:abs_mag2}
M_{i,o} = 0.199[Fe/H] + 0.388.
\end{equation}
\noindent
Similarly, distances in kpc for each red clump star were calculated via the 
relation:

\begin{equation} \label{eq:distance}
d=\frac{10\left [ 10^{\left ( i_{o}-M_{i,o} \right )/5} \right ]}{1000}.
\end{equation}

The derived distance distribution is shown in the right panel of 
Figure \ref{fig:pm_plot} where we indicate the adopted bulge membership limits 
of 6 $<$ d $<$ 12 kpc.  The distribution is slightly skewed toward the near 
end of the bar with a tail to the far end, but this is likely due to a 
combination of the proper motion cleaning preferentially removing foreground 
stars along with viewing angle effects.  We note also that the distribution in 
Figure \ref{fig:pm_plot} peaks at a distance of $\sim$8.2 kpc, which is in 
good agreement with a recent geometric estimate of 8.178 kpc for the Galactic 
Center by \citet{Gravity19}.

\begin{figure}
\includegraphics[width=\columnwidth]{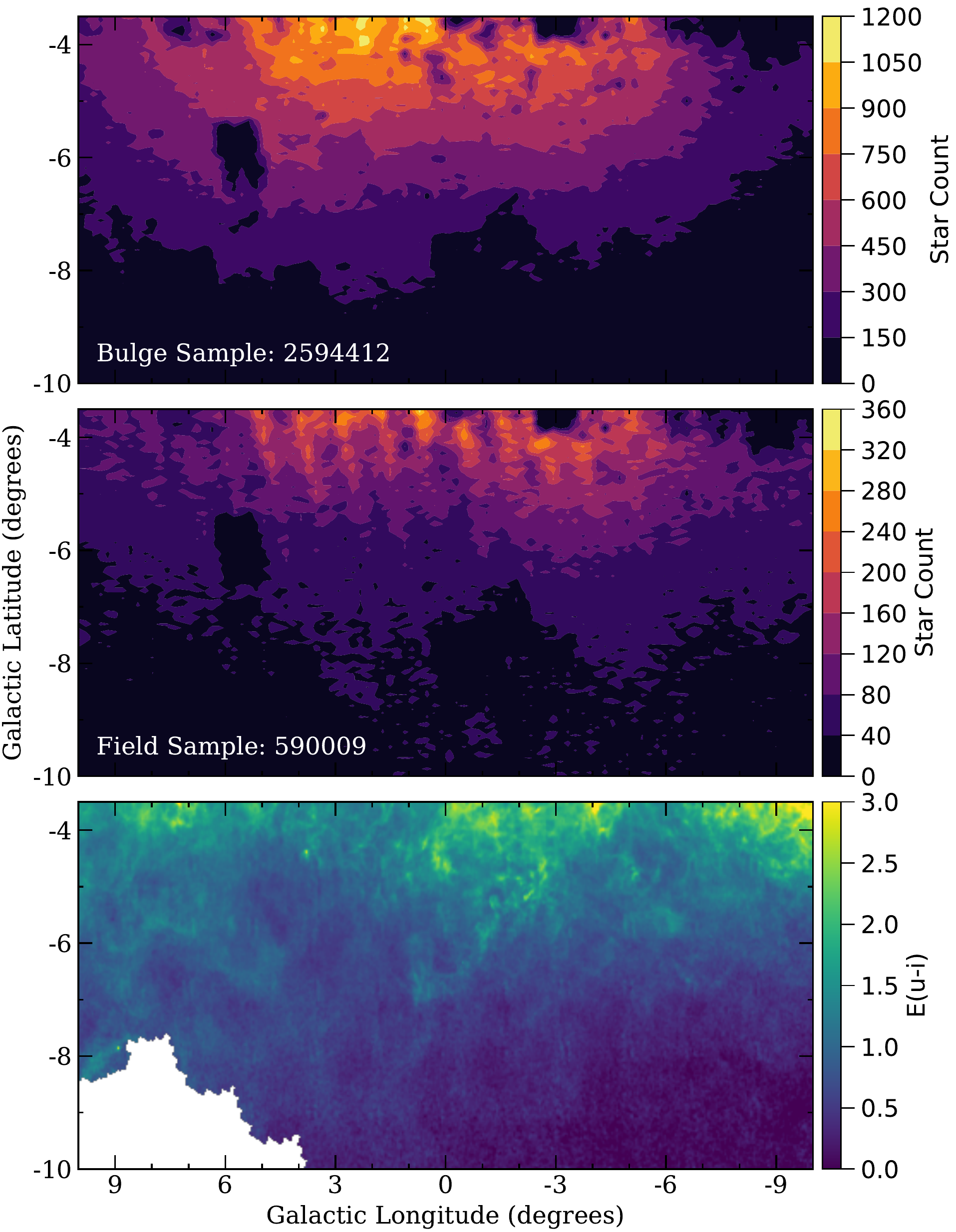}
\caption{Star count contours are shown for our final catalog of likely bulge
red clump stars (top panel) and those rejected as foreground contamination 
(middle panel).  The small ``holes" in the contours represent omitted regions
due to the close proximity (5$\arcmin$) of known globular clusters.  The few
larger low-count regions are due to incomplete sampling in the BDBS data set.
The regular, nearly vertical striping feature, seen especially at positive
longitudes in the field sample and close to the plane, is due to the Gaia 
scanning function.  The bottom panel shows the E($u-i$) reddening adopted 
for this work.}
\label{fig:rc_count_map}
\end{figure}

Figure \ref{fig:rc_count_map} shows the spatial distribution of red clump stars
in our final catalog, along with an E($u-i$) reddening map derived from
\citet{Simion17}.  The catalog reproduces the expected shape of the 
bulge/bar system, and except for a few low count regions due to incomplete
sampling, such as near $(l,b) = (+6^{\circ},-6^{\circ})$, the star counts 
smoothly vary across the BDBS footprint.  The final catalog includes 
$\sim$2.6 million likely red clump bulge members while $\sim$600,000 were 
rejected as likely foreground stars.  A list of coordinates 
(corrected for spatial detector distortions; Clarkson et al., in 
prep.)\footnote{In addition to being corrected for spatial distortions, the 
coordinates are provided in the Gaia EDR3 ICRS frame precessed to the
2014.0 epoch.}, $ugi$ photometry, A$_{\lambda}$, [Fe/H], distance values,
and associated errors is provided in Table \ref{tab:rc_data}.  


\begin{figure}
\includegraphics[width=\columnwidth]{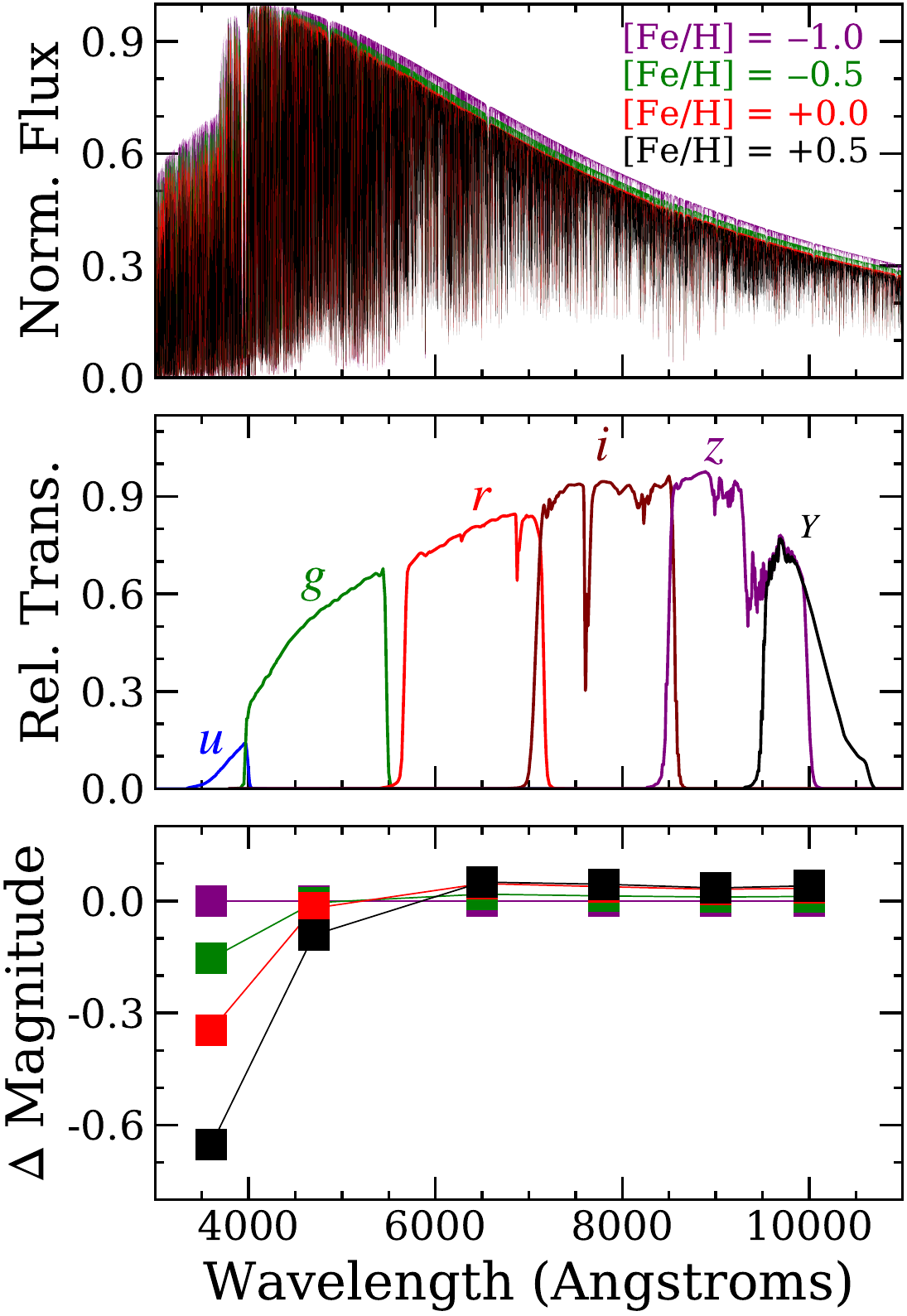}
\caption{\emph{Top:} synthetic high resolution spectra from the POLLUX database
\citep{Palacios10} are shown for typical red clump stars (T$_{eff}$ =
5000 K, log(g) = 2.5 cgs) spanning $-1 <$ [Fe/H] $< +0.5$.  Stars with [Fe/H]
$<$ 0 were modeled with [$\alpha$/Fe] = $+$0.4 while higher metallicity stars
had [$\alpha$/Fe] = 0.  Note that the spectra are normalized to each have the
same peak flux.  \emph{Middle:} filter response curves for the DECam
filters used in BDBS.  \emph{Bottom:} each box shows the difference in
magnitude for a given filter between the [Fe/H] = $-$1 spectrum (purple) and
those with [Fe/H] = $-$0.5 (green), $+$0.0 (red), and $+$0.5 (black).  The
magnitude differences were calculated by convolving the filter response curves
with the original, non-normalized spectra in the top panel.}
\label{fig:spec_compare}
\end{figure}

\section{Red Clump [Fe/H] Measurements from \lowercase{$u$}-band Photometry}
\label{sec:rc_measure}
\subsection{Red Clump Color-Metallicity Relation} \label{ssec:cm_relation}
Near-UV photometry has been used for decades to measure the metallicities of 
stars across a variety of evolutionary states and populations 
\citep[e.g.,][]{Eggen62,Keller07,Ivezic08,Zou16,Ibata17,Howes19,Mohammed19,
Nataf21}.  Figure \ref{fig:spec_compare} highlights the advantage of the 
$u$-band over redder filters when separating red clump stars based on their 
metallicities.  The $u$-band appears to remain sensitive to 
changes in metallicity up to very high values.  Even for [Fe/H] = $+$0.5, 
the cumulative absorption effects of many weaker metal lines blueward of 
$\sim$4000 \AA\ can be seen despite the growing saturation of stronger
features.  This explains the large red clump color dispersions seen in 
Figures \ref{fig:rc_cmd}-\ref{fig:cmd_selection} when the $u$-band is 
included, compared to those with redder bands.

The effects illustrated in Figure \ref{fig:spec_compare} were exploited by 
\citet{Johnson20} and \citet{Lim21} to show that the $(u-g)_{o}$ and $(u-i)_{o}$
colors of BDBS red clump stars are well-correlated with spectroscopic [Fe/H] 
measurements over the full metallicity range of typical bulge stars.  For this
work, we employed the color-metallicity relation given by Equation 22 of 
\citet{Johnson20}:

\begin{equation} \label{eq:cm_calib}
[Fe/H]=0.563(u-i)_{o}-2.074,
\end{equation}
\noindent
which is calibrated with overlapping stars in the GIBS database 
\citep{Zoccali17}.  As mentioned previously, the $u$ and $i$-band photometry, 
extinction coefficients, [Fe/H] values, and associated errors are provided in 
Table \ref{tab:rc_data}.

\begin{figure*}
\includegraphics[width=\textwidth]{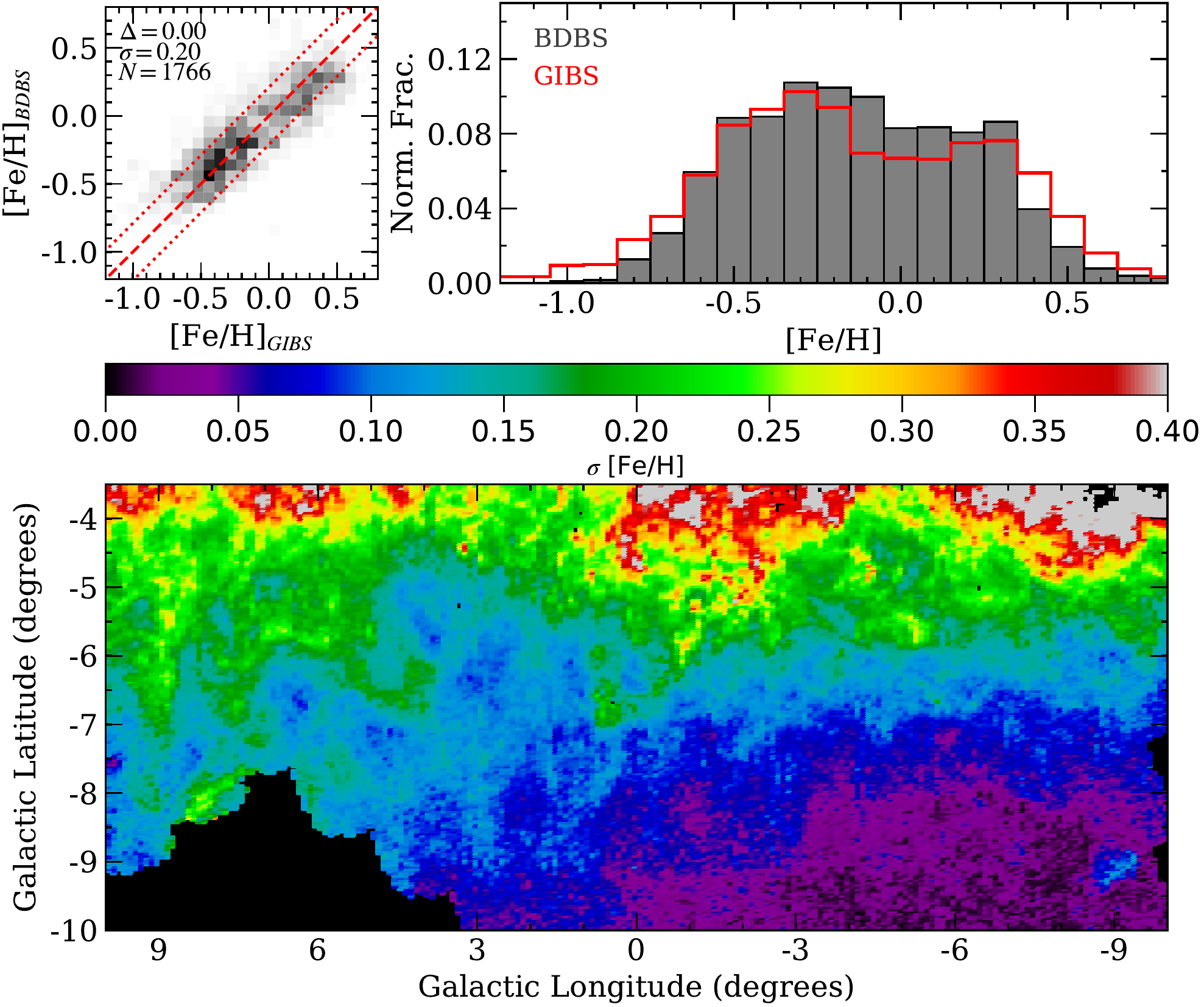}
\caption{\emph{Left:} a density map comparing the photometric (BDBS) and 
spectroscopic (GIBS) [Fe/H] values is shown.  The dashed red line shows the 
fit from Equation \ref{eq:cm_calib} while the dotted red lines show the 
measured 0.2 dex dispersion.  Note that the dispersion value was calculated 
after a single sigma-clipping pass to remove extreme ($>$ 3$\sigma$) outliers. 
The comparison includes $>$ 1700 stars in common between BDBS and GIBS, and 
spans all GIBS fields that overlap with Table \ref{tab:rc_data}.  \emph{Right:}
The grey (BDBS) and red (GIBS) histograms illustrate the derived metallicity 
distribution functions using all stars shown in the left panels.  Both 
distributions are similarly broad and appear bimodal.  \emph{Bottom:} an 
[Fe/H] uncertainty map, based on the errors provided in 
Table \ref{tab:rc_data}, is shown.  Since the $u$-band photometric 
and reddening errors dominate the [Fe/H] uncertainty, this map is 
morphologically similar to the E($u-i$) panel of 
Figure \ref{fig:rc_count_map}.}
\label{fig:feh_comp}
\end{figure*}

The $u$-band photometric and reddening uncertainties are expected to 
dominate the [Fe/H] error budget, along with associated uncertainties
in the transformations between the original VISTA-based extinction map
and our adopted SDSS and Pan-STARRS calibration systems \citep[see 
Section 3.4 of][]{Johnson20}.  The [Fe/H] uncertainty values from 
Table \ref{tab:rc_data} are calculated via error propagation
of Equation \ref{eq:cm_calib}:

\begin{equation} \label{eq:feh_err}
\begin{split}
\sigma_{[Fe/H]}=\Bigl[ \left (\frac{\partial [Fe/H]}{\partial u}\sigma_{u} \right )^{2}+\left (\frac{\partial [Fe/H]}{\partial A_{u}}\sigma_{A_{u}} \right )^{2}+ \\ \left (\frac{\partial [Fe/H]}{\partial i}\sigma_{i} \right )^{2}+\left (\frac{\partial [Fe/H]}{\partial A_{i}}\sigma_{A_{i}} \right )^{2} \Bigr]^{\frac{1}{2}},
\end{split}
\end{equation}

\noindent
where $\sigma_{u}$ and $\sigma_{i}$ are the photometric errors from 
Table \ref{tab:rc_data} and $\sigma_{A_{u}}$ and $\sigma_{A_{i}}$ are 
derived from the E($J-K_{S}$) uncertainty values in the VVV map 
propagated with Equations 14, 15, 16, and 19 of \citet{Johnson20}.
We find a median [Fe/H] uncertainty of 0.19 dex, but the bottom panel of
Figure \ref{fig:feh_comp} shows that the typical uncertainty value is 
correlated with Galactic latitude.  Fields 
with $b < -6^\circ$ generally have $\sigma_{[Fe/H]}$ $<$ 0.15 dex while 
those closer to the plane have uncertainties closer to 0.2-0.3 dex, due 
to the increased crowding and extinction.

However, the top panels of Figure \ref{fig:feh_comp} indicate that the 
BDBS-GIBS calibration is relatively uniform across a variety of sight lines,
with a 1$\sigma$ scatter of 0.2 dex.  We note that 
\citet[][see their Figure 8]{Lim21} found an almost identical calibration and
dispersion when comparing BDBS photometric [Fe/H] values against a separate
spectroscopic metallicity scale published in \citet{Lim21_spec}.

\citet[][see their Figures 18-19]{Johnson20} also showed from comparisons with 
red clump stars in globular clusters that typical BDBS [Fe/H] uncertainties, 
even in very crowded fields, can be as low as $\sim$0.10-0.15 dex across a wide 
range of metallicities.  The top right panel of Figure \ref{fig:feh_comp} also 
indicates that the BDBS-derived metallicity distribution functions are 
generally well-matched against those from GIBS.

Finally, distance uncertainties in Table \ref{tab:rc_data} are calculated
from standard error propagation of Equations \ref{eq:abs_mag2} and
\ref{eq:distance}:

\begin{equation} \label{eq:dist_err}
\sigma_{d}=\sqrt{\left ( \frac{\partial d}{\partial i}\sigma_{i} \right )^{2}+\left ( \frac{\partial d}{\partial A_{i}}\sigma_{A_{i}} \right )^{2}+\left ( \frac{\partial d}{\partial M_{i,o}}\sigma_{M_{i,o}} \right )^{2}},
\end{equation}

\noindent
where the $\sigma_{i}$ and $\sigma_{A_{i}}$ values are the same as those in
Equation \ref{eq:feh_err}.  The $\sigma_{M_{i,o}}$ term is calculated from:

\begin{equation} \label{eq:abs_mag_err}
\sigma_{M_{i,o}}=\sqrt{\left ( \frac{\partial M_{i,o}}{\partial [Fe/H]}\sigma_{[Fe/H]} \right )^{2}+\sigma _{M_{i,o;RC}}^{2}},
\end{equation}

\noindent
where the $\sigma_{[Fe/H]}$ value is taken from Equation \ref{eq:feh_err} and
$\sigma_{M_{i,o;RC}}$ is set to 0.1 mag. in order to roughly account for the
luminosity uncertainty of the red clump phase.

\subsection{Contamination and Biases} \label{ssec:contamination}
Despite our efforts described in Section \ref{sec:data_selection} to isolate a 
clean bulge red clump sample, first ascent RGB overlap and inner disk 
contamination are inevitably still present in the final sample.  Additionally, 
a variety of observational biases exist when observing red clump stars that 
can skew results \citep[e.g., see the discussion in][]{Nataf14}.  Among these 
issues are age variations \citep{Bensby13,Bensby17,Bovy19,Saha19,Hasselquist20},
metal-poor selection effects \citep[e.g.,][]{Nataf14}, and [$\alpha$/Fe] trends
(i.e., whether the bulge [$\alpha$/Fe] ratio flattens out at [Fe/H] $>$ 0 or 
continues to decrease).

\begin{figure}
\includegraphics[width=\columnwidth]{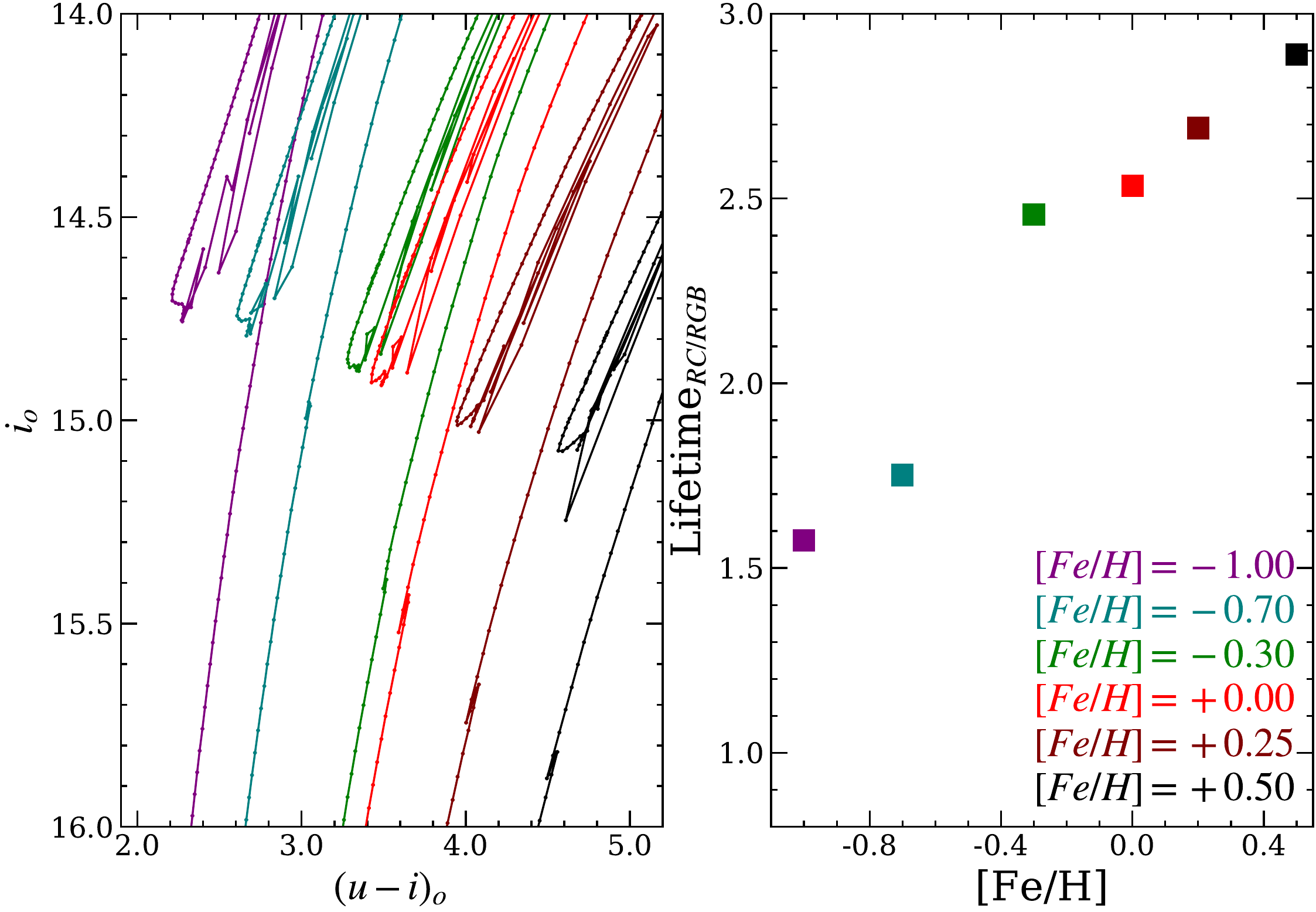}
\caption{\emph{Left:} 10 Gyr MIST isochrones are shown for
[Fe/H] values ranging from $-1$ to $+$0.5, and for which [$\alpha$/Fe] linearly
decreases from $+$0.3 to $+$0.0 between $-0.3 <$ [Fe/H] $< +0.0$.  The
[$\alpha$/Fe] ratios are assumed constant on either side of the [Fe/H] cut-off.
For display purposes, the isochrones have been shifted to a distance of 8 kpc.
\emph{Right:} the relative lifetime ratios of stars passing through RGB and
red clump phases within a typical BDBS selection box are plotted as filled
squares for each track.  The RC/RGB lifetime ratio is a rough tracer of the
expected contamination rate for first ascent RGB stars passing through the
BDBS selection box.}
\label{fig:contamination}
\end{figure}

The bulge exhibits a wide range of [Fe/H] values along all sight lines, and as
a result first ascent RGB stars will overlap with almost any selection 
criteria used to extract red clump stars.  This effect is clearly seen 
in the left panel of Figure \ref{fig:contamination}, which generally covers 
our red clump selection region and shows that red clump and RGB 
stars overlap the same CMD space.  The bar angle depth of $\sim$ 
$\pm$ 2 kpc also adds an additional mixing of evolutionary states.  However, 
the right panel of Figure \ref{fig:contamination} shows that legitimate red 
clump stars still dominate the number density in this region by a factor of 
1.5-3.0, depending on the metallicity.  Since most bulge stars have 
[Fe/H] $>$ $-0.5$, we can reasonably expect that $\sim$70 per cent of the 
targets in our selection boxes are true red clump stars.

\begin{figure}
\includegraphics[width=\columnwidth]{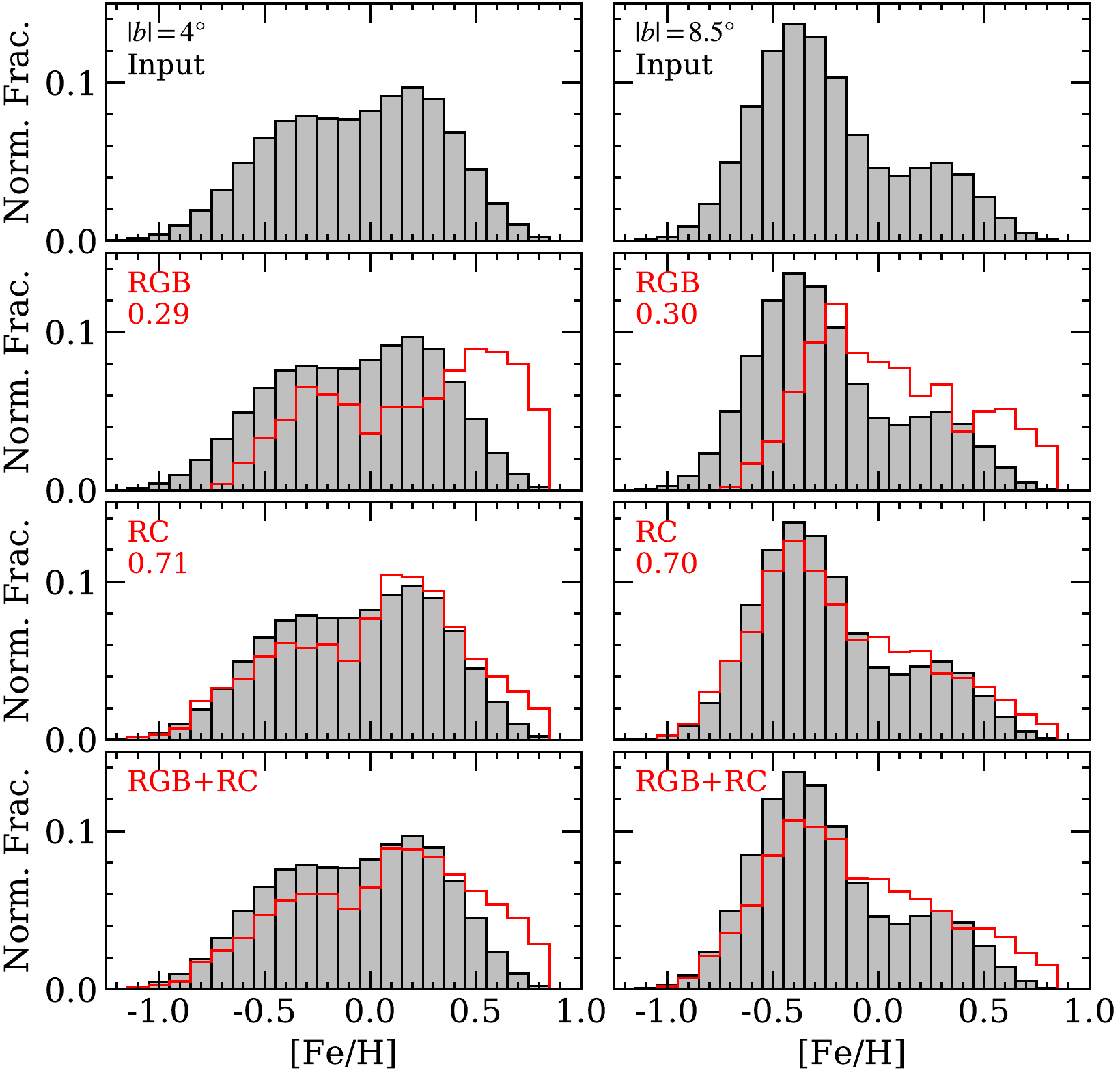}
\caption{\emph{Top row:} input metallicity distribution functions generated from
the two-component Gaussian profile fits of \citet{Zoccali17} for fields with
$|b| = 4^{\circ}$ (inner bulge) and $|b| = 8.5^{\circ}$ (outer bulge).  These
functions are used to test the impact of RGB contamination when deriving 
metallicity distributions from CMD regions dominated by the red clump.
\emph{Second row:} the open red histograms illustrate the photometric 
metallicity distribution functions of RGB stars in each field if the red clump 
color-metallicity relation from Equation \ref{eq:cm_calib} is applied.  
The numbers in each panel are the fractional contributions to expected star 
counts in our adopted red clump selection box.  The grey histograms are the 
original input metallicity distributions from the top row.  
\emph{Third row:} similar to the previous row but for red clump stars.  
\emph{Bottom row:} the combined photometric metallicity distribution functions 
of each field weighted by the expected RC/RGB ratios.}
\label{fig:contamination2}
\end{figure}

We illustrate in Figure \ref{fig:contamination2} the impact on our derived 
metallicity distribution functions when accounting for a mixing of red clump 
stars with those on the RGB.  The top panels illustrate sample metallicity
distributions for inner and outer bulge fields constructed from the analytical
fits given in \citet{Zoccali17}.  Although the \citet{Zoccali17} sample suffers
from many of the same issues noted above, for simplicity we will assume that 
these spectroscopic abundances reflect the true underlying distributions along
the sight lines.  We generated a fine grid ($\Delta$[Fe/H] = 0.01) of 
isochrones by interpolating/extrapolating among the tracks shown in 
Figure \ref{fig:contamination} to construct a mock CMD that reflects the 
expected metallicity distributions.  Stars were placed on the isochrone grid
using the duration of the evolutionary phase that each point occupies as a 
weight.  The middle panels reflect the derived metallicity distribution 
functions of RGB and red clump stars assuming both sets had their $(u-i)_{o}$ 
colors converted to [Fe/H] using Equation \ref{eq:cm_calib} (i.e., assuming 
they are all red clump stars).

The middle panels of Figure \ref{fig:contamination2} show that while the 
red clump sample generally does a reasonable job of recreating the input 
metallicity distribution, the RGB-based distribution is predictably shifted 
towards higher [Fe/H].  The RGB stars also produce an unrealistic metal-rich
tail.  However, the combined distributions in the bottom panels, which reflect
the procedure used in this paper, are morphologically similar to the original
input distributions.  The primary effects of combining RGB and red clump stars
appears to be a reduction in the metal-poor tail, an enhancement in the 
metal-rich tail, and a slight change to the ``gap" distance between the 
metal-poor and metal-rich peaks.

\begin{figure}
\includegraphics[width=\columnwidth]{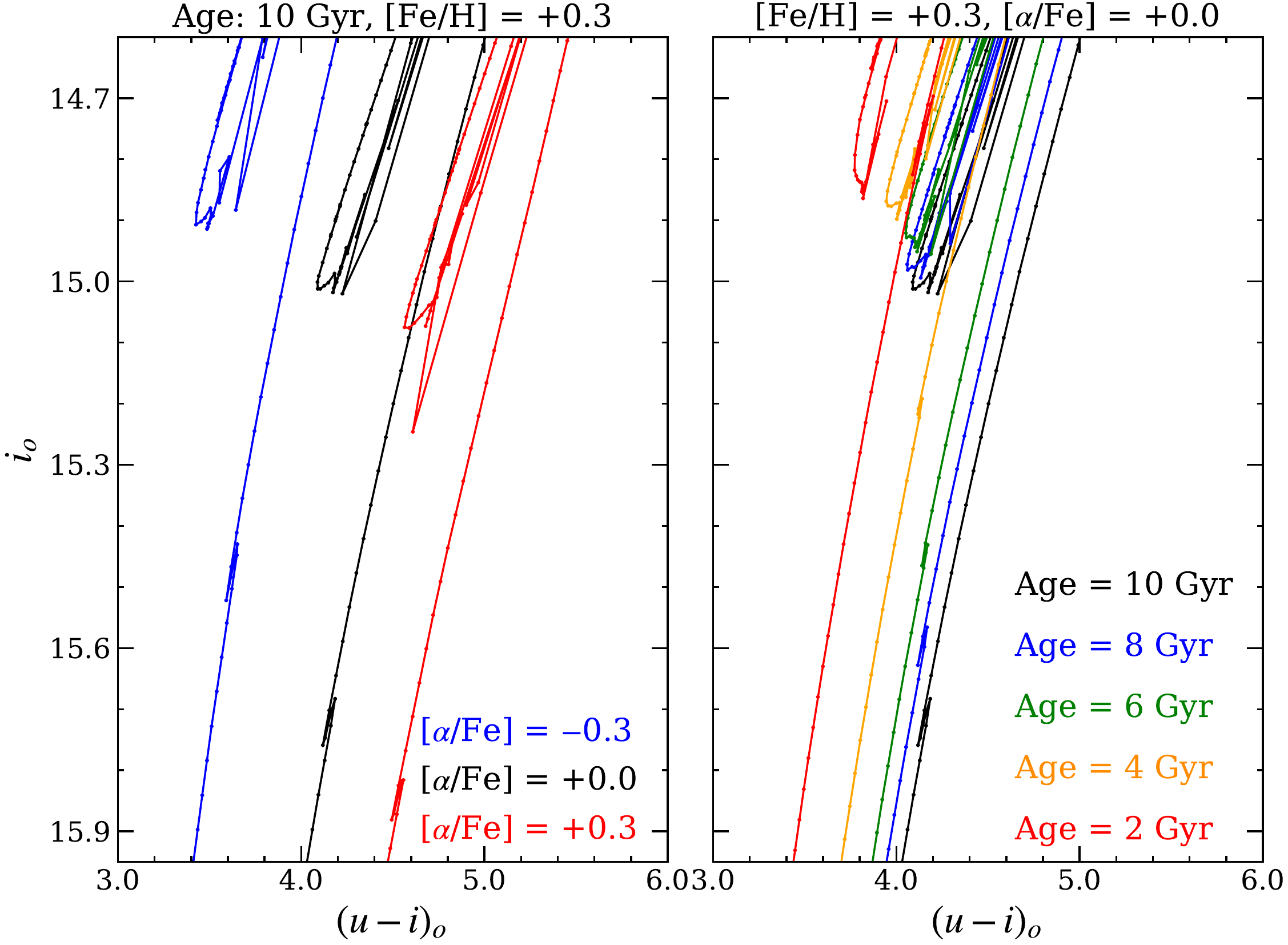}
\caption{The left panel illustrates the isochrone effects of changing the 
[$\alpha$/Fe] composition by $\pm$ 0.3 dex for a 10 Gyr old, metal-rich 
population.  Note that decreasing the [$\alpha$/Fe] composition from 0.0 to
$-$0.3 shifts the red clump $(u-i)_{o}$ distribution blueward by 
$\sim$0.3-0.4 mag.  The right panel shows how age variations can affect the red 
clump location for a typical high metallicity bulge star.  The red clump 
color distribution appears to be mostly insensitive to age variations as long
as stars are older than $\sim$2-4 Gyr.}
\label{fig:age_alpha_effect}
\end{figure}

Age and [$\alpha$/Fe] variations can also affect the $(u-i)_{o}$ distribution 
and thus the derived [Fe/H] values.  Although these variations are likely 
already folded into the empirical color-metallicity relation, it is instructive
to examine how each could affect the derived metallicity distribution 
functions.  We highlight two scenarios in Figure \ref{fig:age_alpha_effect} 
where these effects may manifest.

The left panel of Figure \ref{fig:age_alpha_effect} shows the impact on an
[Fe/H] = $+$0.3, 10 Gyr evolutionary track induced by changing 
[$\alpha$/Fe] $\pm$ 0.3 dex from the expected Solar ratio.  Since modifying 
[$\alpha$/Fe] while holding [Fe/H] constant is equivalent to changing [M/H], 
the isochrones move in the expected direction (i.e., higher metallicity stars 
are redder).  This is potentially relevant to the bulge because: (1) not all 
$\alpha$-elements change in the same way with [Fe/H] 
\citep[e.g.,][]{Fulbright07,Gonzalez11,Johnson14}; and (2) some data show 
[$\alpha$/Fe] leveling off at [Fe/H] $>$ 0 while others 
find that it continues to decrease with increasing metallicity \citep[e.g., see
Figure 6 of][and references therein]{Barbuy18}.  However, for the bulge stars
where this impact would be maximized ([Fe/H] $\ga$ +0.2), 
Figure \ref{fig:age_alpha_effect} suggests that a 0.3 dex uncertainty in 
[$\alpha$/Fe] would only change the measured photometric [Fe/H] by 
$\sim$0.1-0.2 dex.  Fortunately, no investigations have found compelling 
evidence that the mean [$\alpha$/Fe] trends differ significantly between fields.

Effects on the photometric metallicity values due to age variations are more
concerning.  While likely limited to stars with [Fe/H] $\ga$ 0, some evidence
suggests that the prominence of young stars may be a strong function of the
bulge/bar location being probed \citep[e.g.,][]{Ness14,Hasselquist20}.
Although \citet{Bensby13,Bensby17} find evidence supporting the existence of 
bulge stars as young as $\sim$2-3 Gyr, the general consensus is that stars 
significantly younger than 10 Gyr are probably found very close to the plane 
\citep[e.g.,][]{Ness14,Hasselquist20}.  The right panel of 
Figure \ref{fig:age_alpha_effect} shows that even if a young population exists 
in our fields these stars will not change the red clump color distribution in 
any significant way, as long as most stars are older than $\sim$2-4 Gyr.  
This assumption is further supported by recent analyses from 
\citet{Marchetti22} and \citet{Joyce22}, which do not find evidence supporting a
significant population of stars with ages below approximately 6 Gyr.

Finally, we note that disk and halo contamination rates for most bulge fields
should be relatively small.  For example, \citet{Zoccali08} used the 
Besancon Milky Way model \citep{Robin03} to calculate that the thin disk, 
thick disk, and halo likely only contribute about 3, 6, and 0.1 per cent of 
all stars in the red clump region of Baade's window.  Although the disk 
fractions can increase to 10 per cent or more outside $b = -6^{\circ}$, we 
regard these as upper limits since our proper motion cleaning procedure likely 
removed a substantial number of foreground stars.  We further note that because
post-RGB stars with [Fe/H] $<$ $-1$ tend to reside on the blue horizontal 
branch rather than the red clump, our sample is inherently biased against the 
most metal-poor stars in the bulge.  However, while dedicated searches have 
uncovered many metal-poor stars in the bulge \citep[e.g.,][]{Walker91,
GarciaPerez13,Howes14,Howes16,Arentsen20,Savino20,Lucey21}, surveys targeting
the RGB, rather than just the red clump \citep[e.g.,][]{Fulbright06,
Zoccali08,Johnson11,Johnson13_offaxis,Rojas20}, still find that stars with
[Fe/H] $<$ $-1$ contribute only a small fraction ($\la$ 5 per cent) to the
total bulge population.

\section{Results and Discussion} \label{sec:results}
\subsection{BDBS Metallicity Distribution Functions} \label{ssec:mdf}
Numerous surveys have obtained high resolution spectroscopic [Fe/H] abundances
for hundreds to thousands of bulge dwarf, red clump, and/or RGB stars
\citep[e.g., see Section 3.2.1 of][ and references therein]{Barbuy18} and
discovered several persistent trends.  All surveys find that most bulge stars 
are between $-1.0 <$ [Fe/H] $< +0.5$, and that the metallicity
dispersion is large ($\sigma_{[Fe/H]} > 0.3$).  A strong vertical metallicity
gradient is universally observed with the bulge becoming more metal-poor at 
larger distances from the plane.  Additionally, most studies find at least two 
peaks in the metallicity distribution functions that are often separated by 
$>$ 0.3 dex in [Fe/H], and that the amplitudes of the metal-poor and 
metal-rich peaks vary with Galactic latitude.  The metallicity 
distributions are often decomposed into two or more populations using Gaussian 
Mixture Model methods.

\begin{figure}
\includegraphics[width=\columnwidth]{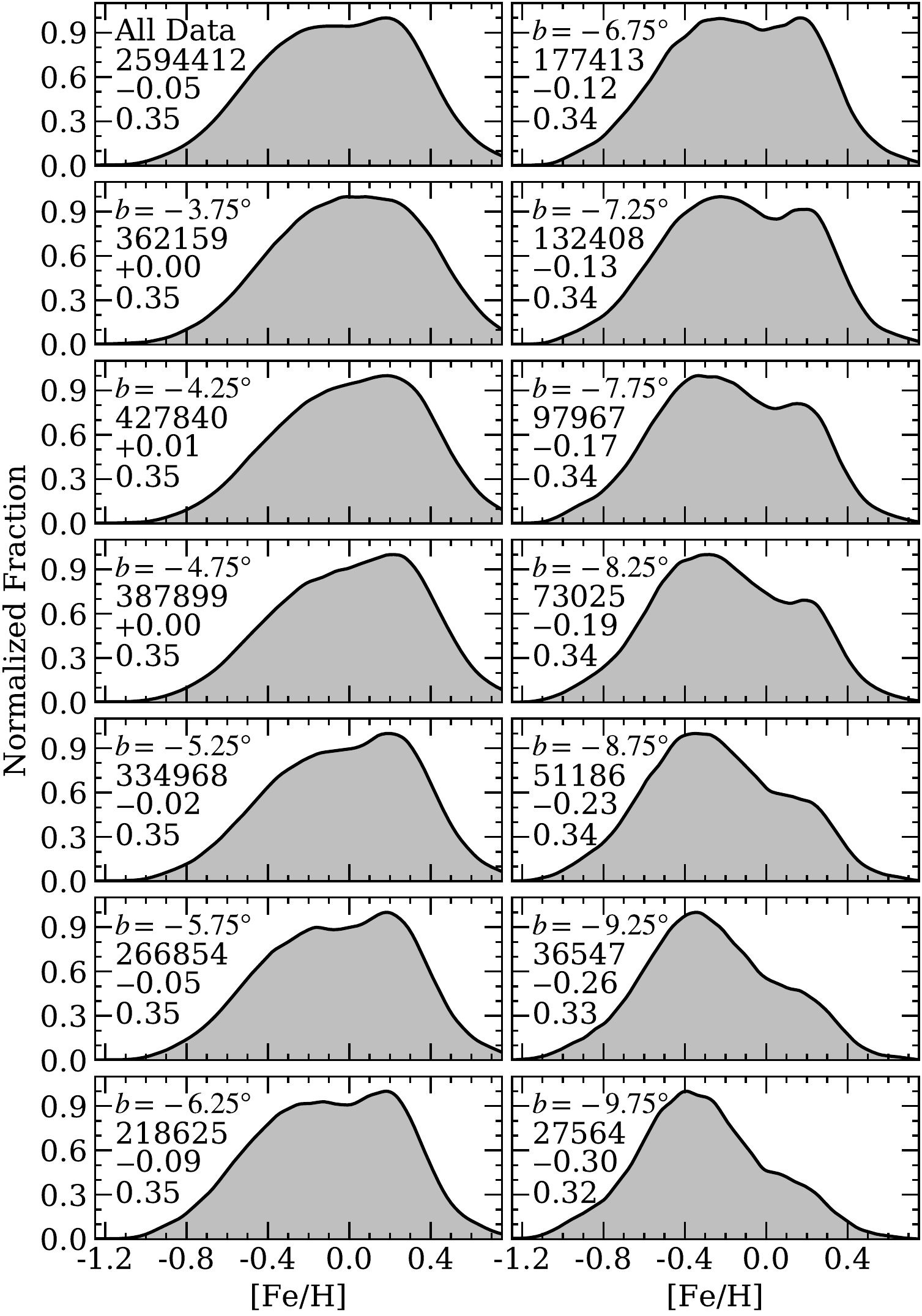}
\caption{The top left panel shows the metallicity distribution function of all
likely bulge red clump stars included in Table \ref{tab:data_table}.
Similarly, the remaining panels show the red clump metallicity distribution
functions summed across all longitudes, using vertical strips spanning $\pm$
0.25$^{\circ}$ in latitude.  The number of stars, center stripe positions,
mean [Fe/H], and $\sigma_{[Fe/H]}$ values are included for all panels.  Note
the strong change in distribution morphology as a function of Galactic
latitude.  Most fields clearly show two distribution peaks.}
\label{fig:mdf_lat}
\end{figure}

\begin{figure}
\includegraphics[width=\columnwidth]{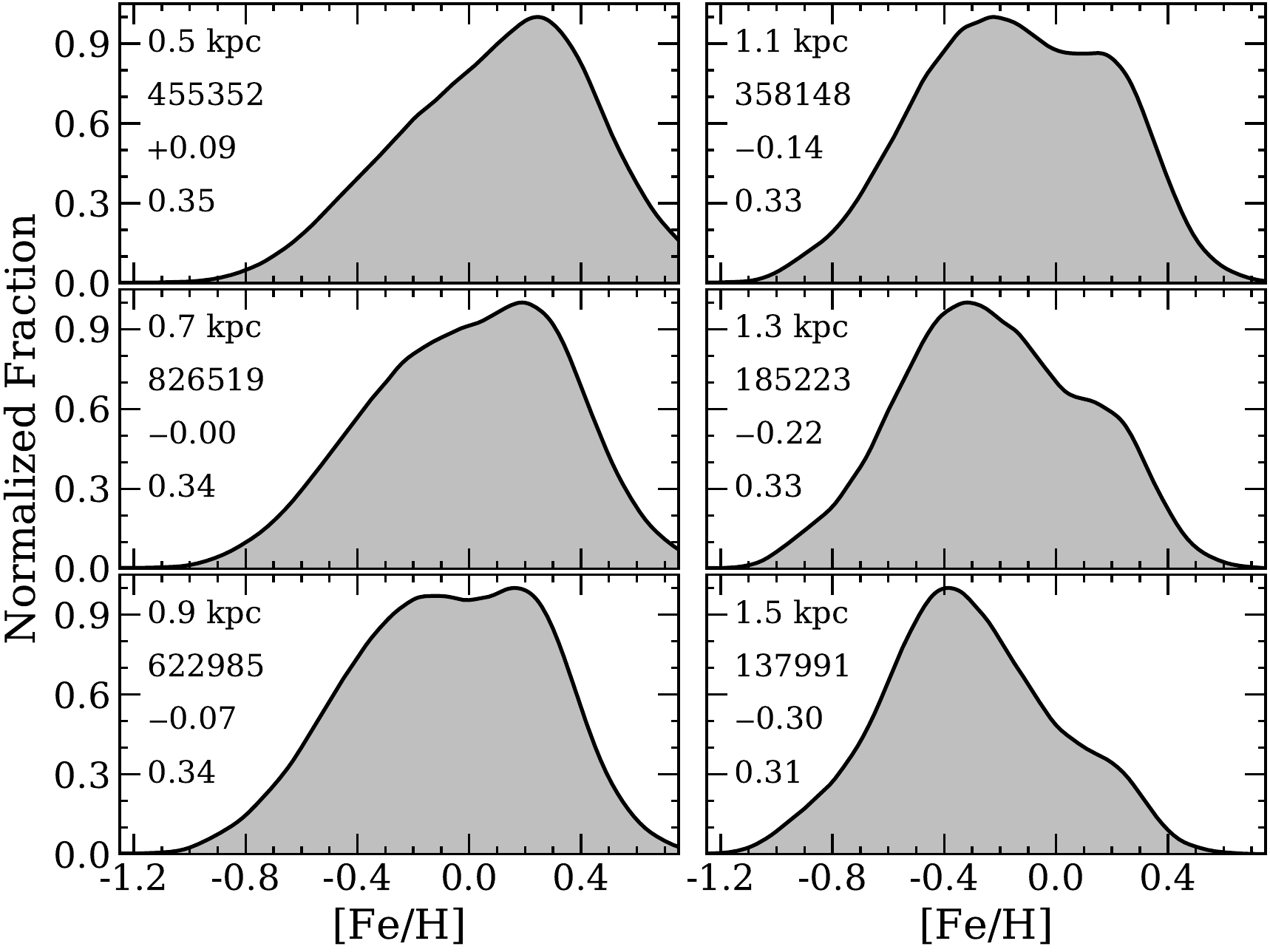}
\caption{Similar to Figure \ref{fig:mdf_lat}, red clump metallicity distribution
functions are shown as functions of distance from the Galactic plane (|Z|).
Each slice encompasses all longitudes but only includes Z distances $\pm$ 0.1 
kpc from the values given in each panel.  The results are qualitatively in
agreement with those of Figure \ref{fig:mdf_lat} but the various peaks are
sharper.}
\label{fig:mdf_z}
\end{figure}

Despite these advances, several important questions remain unanswered:
\begin{itemize}
\item How is the metallicity gradient manifested?  Is it simply the mixture of
two populations with different mean [Fe/H] values that vary in scale height, 
or are vertical gradients intrinsic to one or more of these populations?  
\item Should the bulge instead be viewed as a ``continuum" rather than a 
combination of discrete populations?
\item Is the bulge distinct from the disk, or is it well-described as a 
composite of the thin and thick disks?
\item If adopting a multi-component model for the bulge, are Gaussian 
distributions physically motivated or meaningful?  This is especially important
because authors often disagree on the number of components present, even in the
same fields with largely overlapping samples, along with the individual fit 
widths, centroids, and amplitudes.
\item Do the distribution tails hold any information about the bulge formation
process?
\end{itemize}

Therefore, we add here new metallicity distribution functions for all 2.6 
million red clump stars listed in Table \ref{tab:data_table}.  The BDBS 
data set is at least 100-1000$\times$ larger than previous spectroscopic 
surveys, and includes nearly uniform coverage between $|l| < 10^{\circ}$ and
$-10^{\circ} < b < -3.5^{\circ}$.  Broad summaries of the data
are provided in Figures \ref{fig:mdf_lat}-\ref{fig:mdf_bdz} as summed 
distributions\footnote{For the present work, a ``summed distribution" 
refers to combining the [Fe/H] values for all stars meeting criteria of
interest, such as residing within a given range of Galactic latitudes and/or
longitudes.} across all available longitudes in 0.5$^{\circ}$ latitude and 
0.2 kpc slices, respectively.  We outline below several key observations 
related to changes in the distribution morphology, including: peak locations, 
gradients, and tails.

\subsubsection{Metallicity Distribution Function Morphology} \label{sssec:morph}
\begin{figure*}
\includegraphics[scale=0.7]{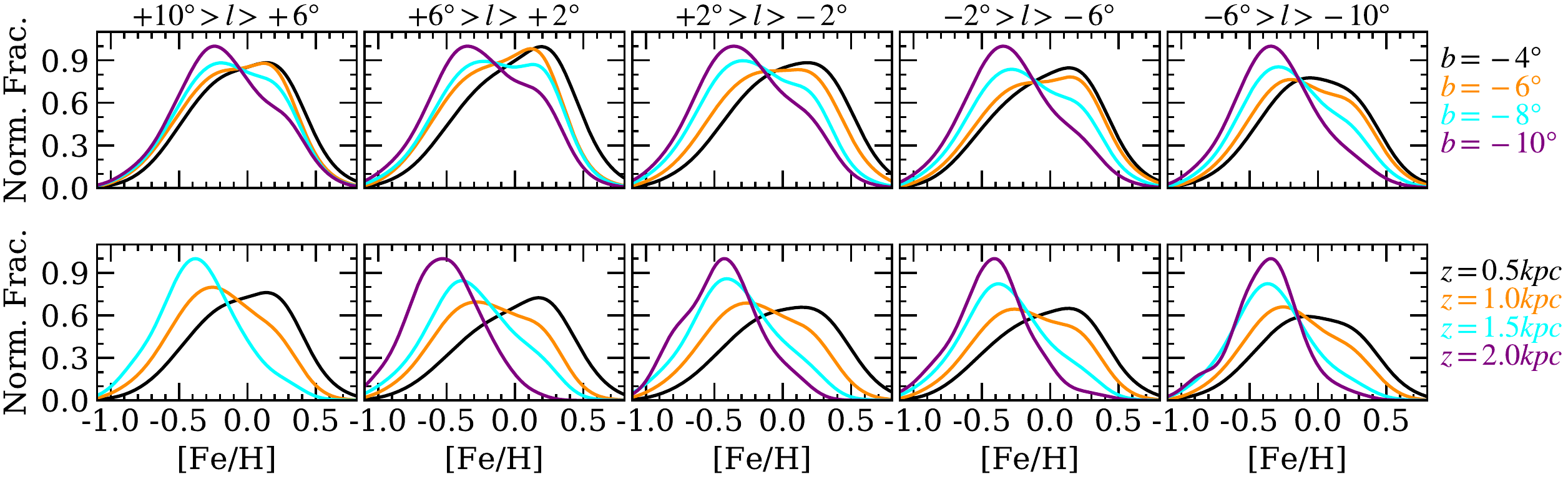}
\caption{The top row shows metallicity distribution functions across 5 sight
lines, each spanning 4 degrees in longitude.  The black, orange, cyan, and
purple lines show the metallicity distributions for two degree increments in
latitude ranging from $b = -4^{\circ}$ to $-10^{\circ}$.  The data are
normalized such that the peak in the coordinate slice containing the most
stars equals unity.  The bottom panels show similar slices except the data
are separated by vertical distance from the plane (|Z|).  The data cover
0.5 kpc slices ranging from 0.5-2.0 kpc.  Note that the left panel does not
have a 2 kpc distribution because observing these stars on the near side of the
bar would require observing at $b < -10^{\circ}$.}
\label{fig:mdf_bdz}
\end{figure*}

Visual inspection of Figures \ref{fig:mdf_lat}-\ref{fig:mdf_bdz} shows that 
while all of the slices contain very broad distributions, the morphological 
shapes are strongly correlated with distance from the Galactic plane.  The 
fields closest to the plane ($b > -5^{\circ}$ and |Z| $<$ 0.6 kpc) are strongly
skewed to peak near [Fe/H]~$\sim$~$+$0.2, and in some cases may be unimodal.  
Comparing the $b = -4.25^{\circ}$ and |Z| = 0.5 kpc panels of 
Figures \ref{fig:mdf_lat}-\ref{fig:mdf_bdz} highlights that 
switching between observed ($b$) and physical (Z) coordinates modifies the 
distribution shapes.  For example, the |Z| = 0.5 kpc panel of 
Figure \ref{fig:mdf_z} shows a more unimodal distribution and narrower
metal-rich peak than the $b = -4.25^{\circ}$ panel of Figure \ref{fig:mdf_lat},
despite a significant overlap in stars.  The reduced metal-poor peak
when working in physical coordinates is likely driven by a removal of stars
that lie along the same projected sight line in latitude but that are at 
farther distances, and thus larger |Z|, where the mean metallicity is lower.

The |Z| = 0.5 kpc distributions in Figure \ref{fig:mdf_bdz} are morphologically
similar to the closed-box enrichment models described in earlier bulge work 
\citep[e.g.,][]{Rich90}.  The exception appears to be the fields located
close to the plane but on the far side of the bar ($-6^{\circ} > l > 
-10^{\circ}$).  Figure \ref{fig:mdf_bdz} shows that these fields are skewed 
toward a lower mean metallicity, and also exhibit a paucity of metal-rich 
stars when compared against similar sight lines on the near side and minor 
axis of the bulge.  A similar behavior was noted by \citet{Zoccali17} in their 
($-8^{\circ},-6^{\circ}$) field, but they did not find any clear explanation
for this discrepancy.  

The remaining fields with |Z| < 1.5 kpc clearly show two peaks centered near 
[Fe/H] $\sim$ $-0.3$ and $+$0.2, but similar to previous work we find that the 
amplitudes change substantially when moving away from the plane.  Fields 
interior to $b \sim -6^{\circ}$ and |Z| $\sim$ 0.8 kpc are skewed such that 
the global maxima are at the metal-rich peaks.  For a narrow strip near 
$b = -6.5^{\circ}$ and |Z| = 1 kpc, the two peaks are equivalent in amplitude. 
Increasing the distance from the plane further leads to a rapid 
decline in the amplitude of the metal-rich local maxima.  

The metallicity distributions in the outer fields ($b < -8.5^{\circ}$ and
|Z| $>$ 1.5 kpc) are dominated by a broad, metal-poor population.  However,
despite the rapid decrease in amplitude of the metal-rich peak, it is still
detectable in all fields with |Z| $<$ 1.75 kpc.  The bottom panels of 
Figure \ref{fig:mdf_bdz} show that the outermost fields in our survey (|Z|
$\sim$ 2 kpc) do not exhibit a clear metal-rich peak, and instead several of
the fields have a secondary, more metal-poor peak near [Fe/H]~$\sim$~$-0.8$.
This population is likely related to the ``C" group identified in 
\citet{Ness13_mdf} as the inner thick disk.  We note that while the 
metallicity distribution changes substantially over a $\sim$ 1.5 kpc vertical
slice of the bulge, the [Fe/H] dispersion remains nearly constant with 
$\sigma_{[Fe/H]} \sim 0.35$.

\begin{figure*}
\includegraphics[width=\textwidth]{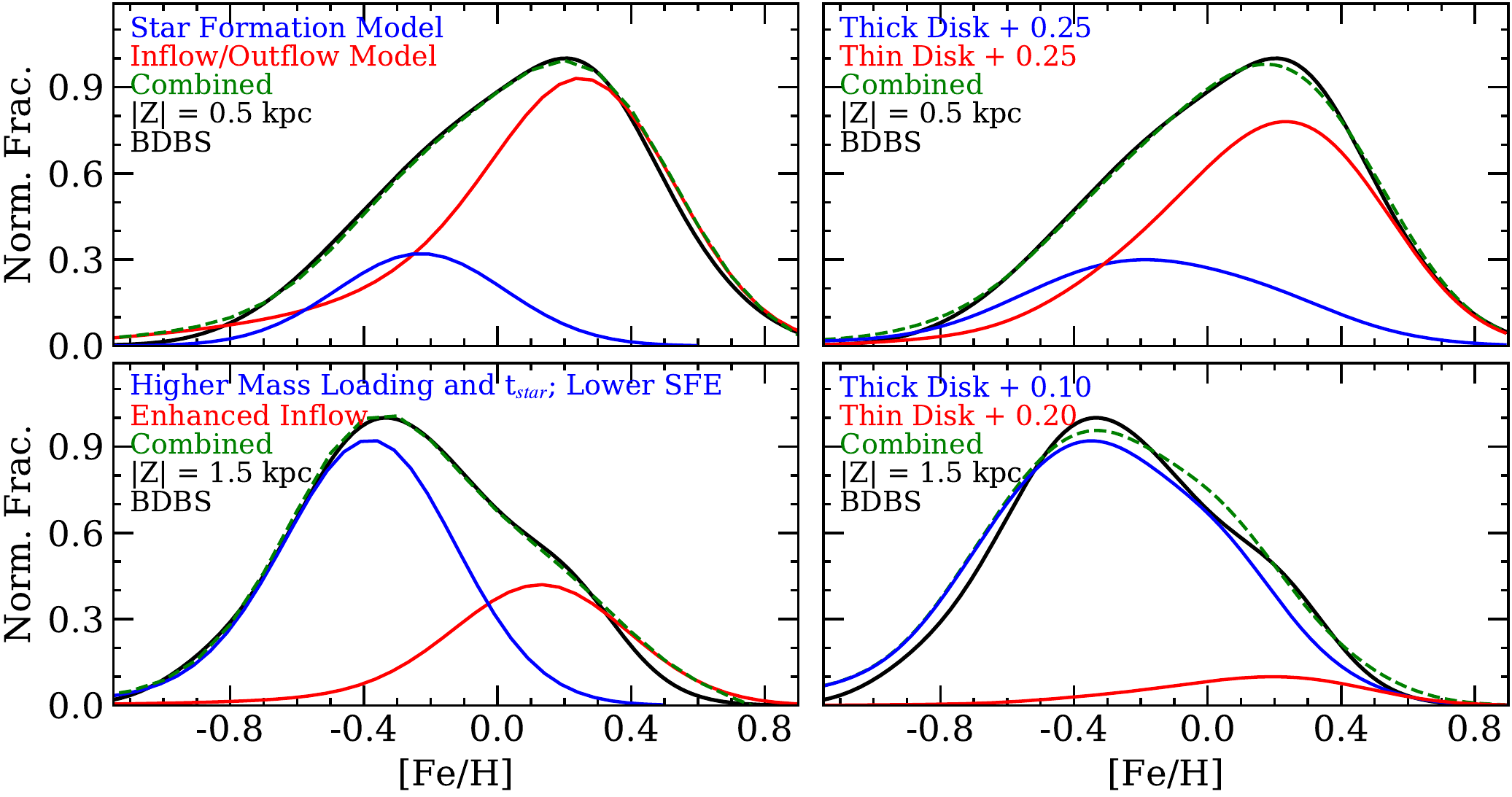}
\caption{The top and bottom rows show the observed BDBS metallicity 
distribution functions (solid black lines) for bulge stars with 
0.3 $<$ |Z| $<$ 0.7 kpc and 1.2 $<$ |Z| $<$ 1.8 kpc, respectively.  
The left panels compare the observed distributions against two analytical 
chemical enrichment models generated with the One-zone Model for the 
Evolution of GAlaxies (OMEGA) code \citep{Cote17}.  For both left panels, 
the blue and red lines represent the ``star formation" and ``inflow/outflow" 
OMEGA models that roughly correspond to the metal-poor and metal-rich 
components from \citet{Grieco12}.  The sum of the two distributions are 
shown as dashed green lines, and all model outputs have been broadened by 
0.25 dex to account for observational uncertainty.  Compared to the top left 
models, the bottom left star formation model has an enhanced mass loading 
factor, a longer star formation time scale (t$_{star}$), and a lower star 
formation efficiency (SFE) to reduce the mean 
yield and broaden the metal-poor tail.  Similarly, the inflow/outflow model
has enhanced inflow of metal-poor gas to reduce the metal-rich tail.  The 
right panels compare the BDBS distributions against the empirical thick 
(red) and thin (blue) disk distributions provided by \citet{Bensby14}.  
Both disk populations have been shifted upward by 0.10-0.25 
dex to better align with the observed distribution peaks.  The thin and thick
disk distributions have also been broadened by 0.20 dex.}
\label{fig:chem_models}
\end{figure*}

\subsubsection{Metallicity Distribution Models} \label{sssec:mdf_models}
Examining the most interior and exterior fields of
Figures \ref{fig:mdf_z}-\ref{fig:mdf_bdz}, which should be dominated by the
distributions that produce the metal-poor and metal-rich peaks, indicates that
neither is likely well-described by a simple convolution of Gaussian models.  
Instead, these distributions appear to have broad and often asymmetric tails 
that likely hold important information about the bulge's star formation history.

\begin{figure*}
\includegraphics[width=\textwidth]{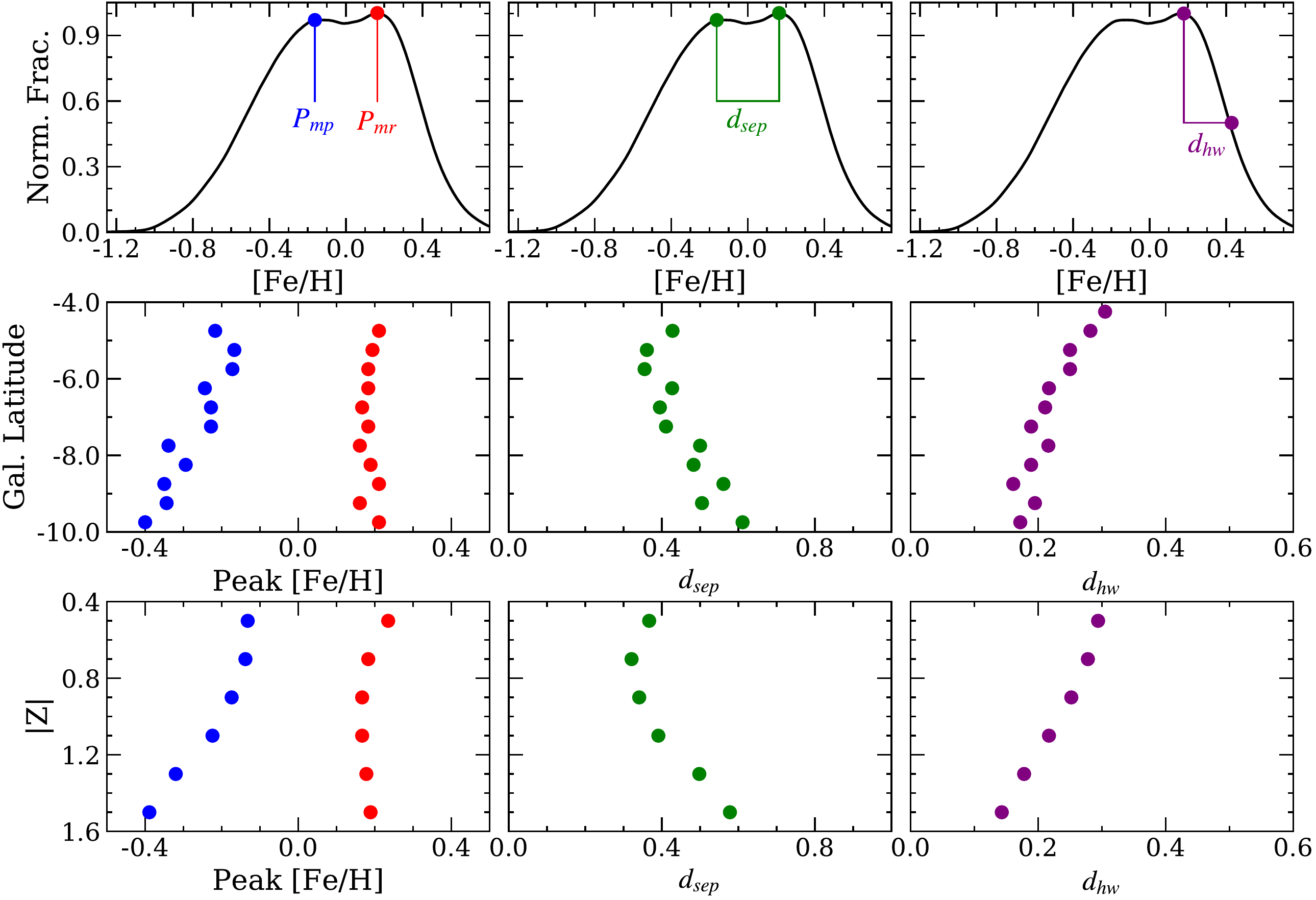}
\caption{\emph{Top:} An example metallicity distribution function is shown
to highlight the various components that are measured for the lower panels.
The top left panel shows how the metal-poor ($P_{mp}$) and metal-rich
($P_{mr}$) peak locations are defined, the middle panel illustrates the
distance between peaks ($d_{sep}$), and the right panel indicates the
distance between the metal-rich peak and the [Fe/H] value at half power
($d_{hw}$).  \emph{Middle:} the various metrics are measured using the
distributions shown in Figure \ref{fig:mdf_lat}.  Note that the metal-rich
peak remains roughly constant while the metal-poor peak becomes progressively 
more metal-poor with increasing distance from the plane.  Similarly, the 
$d_{sep}$ parameter also becomes larger with vertical distance.  The right panel
shows that the metal-rich tail becomes more narrow farther from the plane.
\emph{Bottom:} Identical to the middle panels except measured using
Figure \ref{fig:mdf_z}.}
\label{fig:mdf_decomp}
\end{figure*}

The left panels of Figure \ref{fig:chem_models} compare two sets of chemical 
enrichment models generated with the One-zone Model for the Evolution of 
GAlaxies \citep[OMEGA;][]{Cote17} code against the observed BDBS
metallicity distribution functions derived from stars residing in horizontal
stripes centered near |Z| = 0.5 (``inner bulge") and 1.5 kpc (``outer bulge").
For both models, the inflow gas is assumed to have [Fe/H]~=~$-$1.3 for the 
metal-poor (blue) model and [Fe/H]~=~$-$0.5 for the metal-rich (red) model.  
The metal-poor and metal-rich components of the inner bulge models roughly 
mimic those presented in \citet{Grieco12} that have rapid ($<$ 0.3 Gyr)
and prolonged ($\sim$ 3 Gyr) star formation time scales, 
respectively\footnote{Note that the models presented here are only for 
qualitative comparison against the observed data and have not been optimized 
against parameters such as expected [$\alpha$/Fe] distributions, Galactic 
star formation rates, etc.}.  

Although a detailed modeling of the bulge enrichment history is beyond the 
scope of this paper, the simple models shown in Figure \ref{fig:chem_models}
highlight the possible information content that may be extracted from the 
distribution morphology, particularly the tails.  Changes to the observed 
metallicity distribution functions that correlate with distance from the 
plane can be modeled, at least in part, by changing the model
inflow/outflow rates, star formation time scales/efficiencies, and/or mass 
loading factors ($\eta =\frac{\dot{M}_{outflow}}{\dot{M}_{SFR}}$).  For the 
metal-poor component, a combination of higher mass loading factor, longer star 
formation time scale, and lower star formation efficiency can reduce the 
effective yield from [Fe/H] = $-0.2$ to $-0.4$ while simultaneously broadening 
the metal-poor tail.  Similarly, enhanced inflow of metal-poor gas helps to 
truncate the high metallicity tail of the metal-rich component, though an 
increase in the star formation rate, or the adjustment of a similar lever, 
is required to maintain the high effective yield.  More detailed modeling can 
be obtained by better constraining these free parameters using the abundance 
ratios of elements other than Fe.

Nevertheless, the modifications required to match the inner bulge models to 
the outer bulge data suggest that Galactic winds and related phenomena may have 
played a strong, and likely detectable, role in shaping the bulge's star 
formation history.  Consequently, Figure \ref{fig:chem_models} also 
highlights that the detailed morphologies of the underlying distributions 
can have a dramatic impact on the assumed contribution from each population.
For example, several authors decompose bulge metallicity distributions using
Gaussian mixture models \citep[e.g.,][]{Ness13_mdf,Zoccali17,Rojas20}, but
these models generally have little physical motivation and do not account for a 
population's likely star formation history.  A contrast is seen for
the metal-rich component's shape when comparing Figure \ref{fig:chem_models}
to Figure 7 of \citet{Zoccali17}.  The former model contributes stars from the
metal-rich component down to [Fe/H] $<$ $-0.5$ while the latter fit contributes 
no stars below [Fe/H] $\sim$ $-0.2$.  Additionally, [$\alpha$/Fe] versus 
[Fe/H] plots of bulge stars \citep[e.g., Figure 1 of ][]{Griffith21} show that 
the $\alpha$-enhanced stars span at least [Fe/H] = $-1.5$ to $+$0.3 while the
low-$\alpha$ stars span at least [Fe/H] = $-0.5$ to $+$0.5, which is more
in agreement with the models shown in Figure \ref{fig:chem_models} and 
\citet{Grieco12}.

\begin{figure}
\includegraphics[width=\columnwidth]{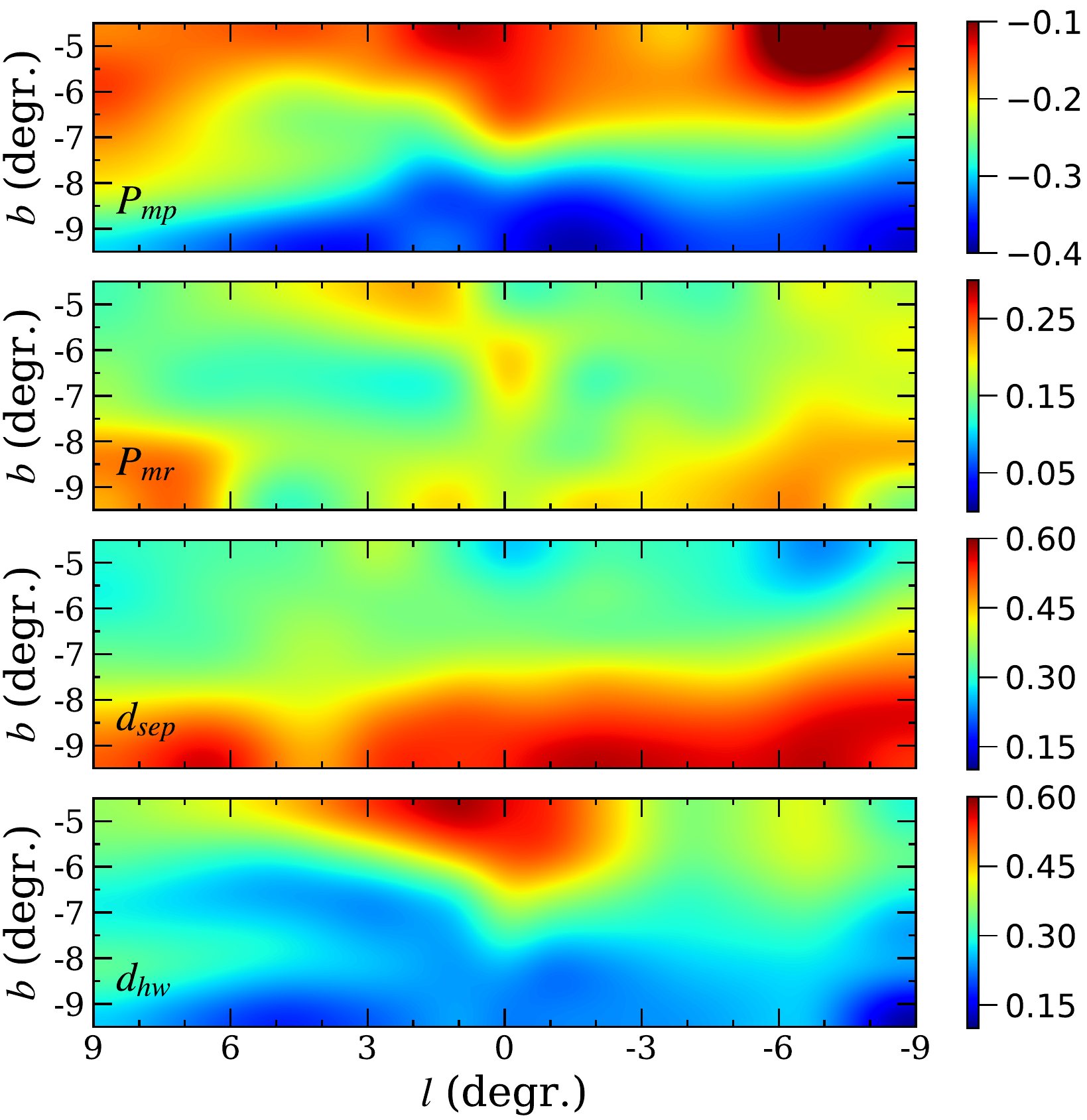}
\caption{From top to bottom the panels show interpolated maps illustrating the 
2D dependence of the metal-poor peak position ($P_{mp}$), metal-rich peak 
position ($P_{mr}$), separation between the two peaks ($d_{sep}$), and the 
distance between the metal-rich peak and half-power position of the metal-rich 
tail ($d_{hw}$), respectively.  The color bars are indicative of the [Fe/H]
values for each panel (see Figure \ref{fig:mdf_decomp} for definitions of each
metric).  Note the latitude dependence of the $P_{mp}$ and $d_{sep}$
metrics.  The $d_{hw}$ metric generally decreases with increasing distance 
from the plane (i.e., the metal-rich tail becomes shorter), but some 
longitudinal dependence is observed for $b > -6^{\circ}$.  The metric values
were determined in grid points anchored at $l$ = $-$9$^{\circ}$, 
$-$7$^{\circ}$, $-$5$^{\circ}$, $-$3$^{\circ}$, $-$1$^{\circ}$, $+$0$^{\circ}$,
$+$1$^{\circ}$, $+$3$^{\circ}$, $+$5$^{\circ}$, $+$7$^{\circ}$, and 
$+$9$^{\circ}$ along with $b$ = $-4.5$$^{\circ}$, $-5.5$$^{\circ}$, 
$-6.5$$^{\circ}$, $-7.5$$^{\circ}$, $-8.5$$^{\circ}$, and $-9.5^{\circ}$.}
\label{fig:decomp_map}
\end{figure}

Similar to the computed models, Figure \ref{fig:chem_models} shows that the 
empirical thin and thick disk data from \citet{Bensby14} are also asymmetric 
and possess long tails.  These combined distribution functions 
match many of the observed features in the BDBS data, including the general 
shapes of the metal-poor and metal-rich tails, which follows previous claims 
noting possible chemical connections between the local disk and bulge 
\citep[e.g.,][]{Melendez08,AlvesBrito10,Gonzalez11,Jonsson17,Zasowski19}.  
However, the empirical thin and thick disk distributions must be shifted 
by $+$0.10 to $+$0.25 dex in [Fe/H] in order to align with the local maxima 
observed in the bulge distributions.  A possible justification for these
arbitrary shifts is that the bulge enriched more quickly than the local 
disk \citep[e.g.,][]{Zoccali06,Fulbright07,Johnson11,Bensby13,Bensby17,
Rojas17,Duong19_chem,Lucertini21}, and as a result formed from gas that 
was polluted by material with a higher effective yield.  Although the data 
presented here do not necessarily indicate that the bulge is an inner 
extension of the disk, Figure \ref{fig:chem_models} shows that the bulge may 
be well-described by a weighted convolution of two enrichment models that 
possess properties similar to the thin and thick disks.

\subsubsection{Peak Locations} \label{sssec:peaks}
As noted in Section \ref{sssec:morph}, every field except perhaps those 
closest to the plane contain two peaks separated by several tenths of a dex
in [Fe/H].  Figure \ref{fig:mdf_decomp} illustrates our measurement method and 
results for tracking the peak [Fe/H] values for the two maxima based on an
examination of Figures \ref{fig:mdf_lat} and \ref{fig:mdf_z}.  The middle 
left panel of Figure \ref{fig:mdf_decomp} shows that despite a large variation
in amplitude across our fields, the metal-rich centroid position remains 
roughly constant at [Fe/H]~$\sim$~$+$0.2.  The second panel of 
Figure \ref{fig:decomp_map} further emphasizes the general insensitivity of
the metal-rich peak position with bulge location as no strong radial 
(longitudinal) nor vertical gradients are found.  However, 
Figure \ref{fig:decomp_map} does indicate a small number of regions where
the metal-rich peak position may be higher than expected, such as along
the minor axis with $b$ > $-6^{\circ}$ and near ($l,b$) $\sim$ 
($-6^{\circ},-8.5^{\circ}$).

In contrast, the metal-poor centroid position is a strong function of Galactic 
latitude.  Figures \ref{fig:mdf_decomp}-\ref{fig:decomp_map} show that the 
metal-poor centroid changes from [Fe/H] $\sim -0.15$ at $b \sim -5^{\circ}$ to 
[Fe/H] $\sim -0.4$ at $b = -9^{\circ}$, in agreement with previous work by
\citet{Rojas14} but contrasting with the seemingly constant metal-poor peak
position found by \citet[][see their Figure 6]{Zoccali17}\footnote{We note that
\citet{Zoccali17} claim the metal-poor centroids are insensitive to latitude,
but a linear fit to the values given in their Table 3 produces a noticeable
negative slope, indicating a possible gradient in the metal-poor population.}.  
Although the metal-poor peak is not as prominent for fields inside |Z| 
$\sim$ 0.8 kpc in Figure \ref{fig:mdf_z}, we find an almost identical behavior 
when examining centroid locations as a function of physical distance from the 
plane.  

Figure \ref{fig:decomp_map} does not show a significant 
longitudinal dependence for $P_{mp}$ in fields with $b > -7^{\circ}$.  However, 
the higher latitude fields show a mild longitudinal correlation such that in 
sight lines of constant latitude those with positive longitudes have their 
$P_{mp}$ shifted to higher metallicities than those with negative longitudes.  
Since the positive longitude fields are dominated by stars on the near side of 
the bar, it seems likely that the longitudinal gradient in the outer bulge is 
primarily due to our viewing angle of the bulge/bar system.

Combining the metal-poor and metal-rich trends in the middle panels of 
Figure \ref{fig:mdf_decomp} illustrates that the separation between the 
metal-poor and metal-rich peaks becomes larger when moving away from the plane.
The third panel of Figure \ref{fig:decomp_map} also shows a clearly defined
vertical dependence of the $d_{sep}$ metric on latitude along with the same
longitudinal dependence as found for the $P_{mp}$ metric.  
Figure \ref{fig:decomp_map} therefore indicates that the vertical gradient in 
$d_{sep}$ is driven almost entirely by the shifting position of the 
metal-poor peak that correlates with distance from the plane.  This 
observation suggests that an intrinsic metallicity gradient exists in the
bulge, and that the mean metallicity gradient is not driven solely by the 
mixing of ``static" metal-poor and  metal-rich populations in different
proportions with latitude.  

The observed correlation between Z and $P_{mp}$ may be evidence of large-scale
events, such as Galactic outflows/winds or inside-out formation, which can 
shift the metallicity distribution function peak to lower [Fe/H] values at
larger distances from the plane \citep[e.g.,][]{Boeche14,Schlesinger14,
Schonrich17,Nandakumar20}.  Although compelling evidence suggests that 
disrupted globular clusters have played some role in building up the 
inner Galaxy stellar populations \citep[e.g.,][]{Ferraro09,Origlia11,
Schiavon17,Lee19,Ferraro21}, it is unlikely that such a mechanism would 
produce the observed metal-poor gradient.

\subsubsection{Mean Metallicity Gradient} \label{sssec:gradients}
The mean metallicities provided in the panels of Figures \ref{fig:mdf_lat}
and \ref{fig:mdf_z} indicate that a strong vertical metallicity gradient is 
present in the bulge.  In particular, the mean [Fe/H] value decreases from 
$\sim$ $+$0.1 near the plane to $-0.3$ in fields $>$ 1.5 kpc from the plane.  
However, we find no significant differences in mean [Fe/H] values for the 
three most interior fields of Figure \ref{fig:mdf_lat}, which suggests that
the metallicity gradient may be significantly flatter for $|b| < -5^{\circ}$.  
Additionally, we find that the dominant composition gradient is in the 
vertical, rather than radial, direction (see also Section \ref{ssec:zdist}).

Using the values presented in Figure \ref{fig:mdf_lat}, we measure a 
metallicity gradient between $b = -3.5^{\circ}$ and $-10^{\circ}$ of 
d[Fe/H]/d$b$ = $-0.05$ dex degree$^{-1}$.  This is equivalent to a 
metallicity gradient of d[Fe/H]/dZ = $-0.38$ dex kpc$^{-1}$ in physical
coordinates for fields with 0.5 < |Z| < 1.5 kpc.  Our measured 
mean metallicity gradient values are almost identical to those obtained when 
fitting a linear function to the changing peak [Fe/H] values of the metal-poor 
population shown in Figure \ref{fig:mdf_decomp}.  Specifically, we find the 
gradient in the metal-poor peak centroid location to be d[Fe/H]/d$b$ = $-0.04$ 
dex degree$^{-1}$, which is equivalent to d[Fe/H]/dZ = $-0.37$ dex kpc$^{-1}$ 
in physical dimensions.

\subsubsection{Metal-rich Tail Variations} \label{sssec:tail_variations}
A more subtle change in the metallicity distribution function morphology
can be seen upon close inspection of the metal-rich tails in 
Figures \ref{fig:mdf_lat} and \ref{fig:mdf_z}.  Although the metal-rich
peak centroids remain stable across all slices, the distribution shape at
higher metallicities appears to become truncated when examining higher 
latitude fields.  

We investigated the potential metal-rich tail variations using the right panels
of Figure \ref{fig:mdf_decomp}, which measure the distance between the 
metal-rich peak centroid and the [Fe/H] value corresponding to half the peak 
amplitude ($d_{hw}$).  Figure \ref{fig:mdf_decomp} indicates that 
the metal-rich tail forms a progressively narrower distribution at larger 
distances from the plane.  In fact, the $d_{hw}$ parameter exhibits a gradient 
of d($d_{hw}$)/d$b$ = $-$0.02 dex degree$^{-1}$, which is equivalent to 
d($d_{hw}$)/d$Z$ = $-$0.16 dex kpc$^{-1}$.  It is possible that differences in 
star formation efficiency, infall times, and/or outflow may be driving the 
varying metal-rich tail widths as functions of vertical distance 
\citep[e.g.,][]{Mould84,Ballero07}.  Similarly, sudden gas removal that 
becomes more efficient farther from the plane could drive a sharper metal-rich 
cut-off at higher latitudes, as is observed in dwarf spheroidal galaxies
\citep[e.g.,][]{Koch07,Kirby11}.

Figure \ref{fig:decomp_map} confirms that $d_{hw}$ generally decreases with
increasing distance from the plane, although some mild longitudinal dependence
may also be present for fields outside $b = -6^{\circ}$.  A 
much stronger longitudinal dependence is found for fields with $b > -6^{\circ}$.
Sight lines near the minor axis ($|l| < 3^{\circ}$) have noticeably longer
metal-rich tails than fields with $|l| > 3^{\circ}$, perhaps as a consequence
of inside-out formation of the early bulge.  This observation is further 
supported by evidence presented in \citet{Sit20} that for fields close to
the plane the metallicity dispersion reaches a maximum near the Galactic 
Center but decreases when moving outward along the major axis.  We note also
that the region of maximum $d_{hw}$ in Figure \ref{fig:decomp_map} also
coincides with a population of stars having noticeably higher 
$\sigma_{\mu_{b}}$ compared to those in adjacent fields \citep[see Figure 10 
of ][]{Clarke19}.

We caution that the $d_{hw}$ metric could be particularly sensitive to RGB 
contamination in the red clump region.  The contamination simulations presented
in Figure \ref{fig:contamination2} do produce broader metal-rich tails when RGB
contamination is accounted for.  Alternatively, the length of the metal-rich 
tail could be sensitive to reddening uncertainties, with weaker tails appearing 
at higher latitudes due to intrinsically reduced differential extinction (see
also the bottom panel of Figure \ref{fig:feh_comp}).  However, as noted 
previously the [Fe/H] dispersion for the metallicity distribution in a given 
field is only weakly dependent on latitude.

\subsubsection{Literature Comparisons} \label{sssec:literature}
\begin{figure}
\includegraphics[width=\columnwidth]{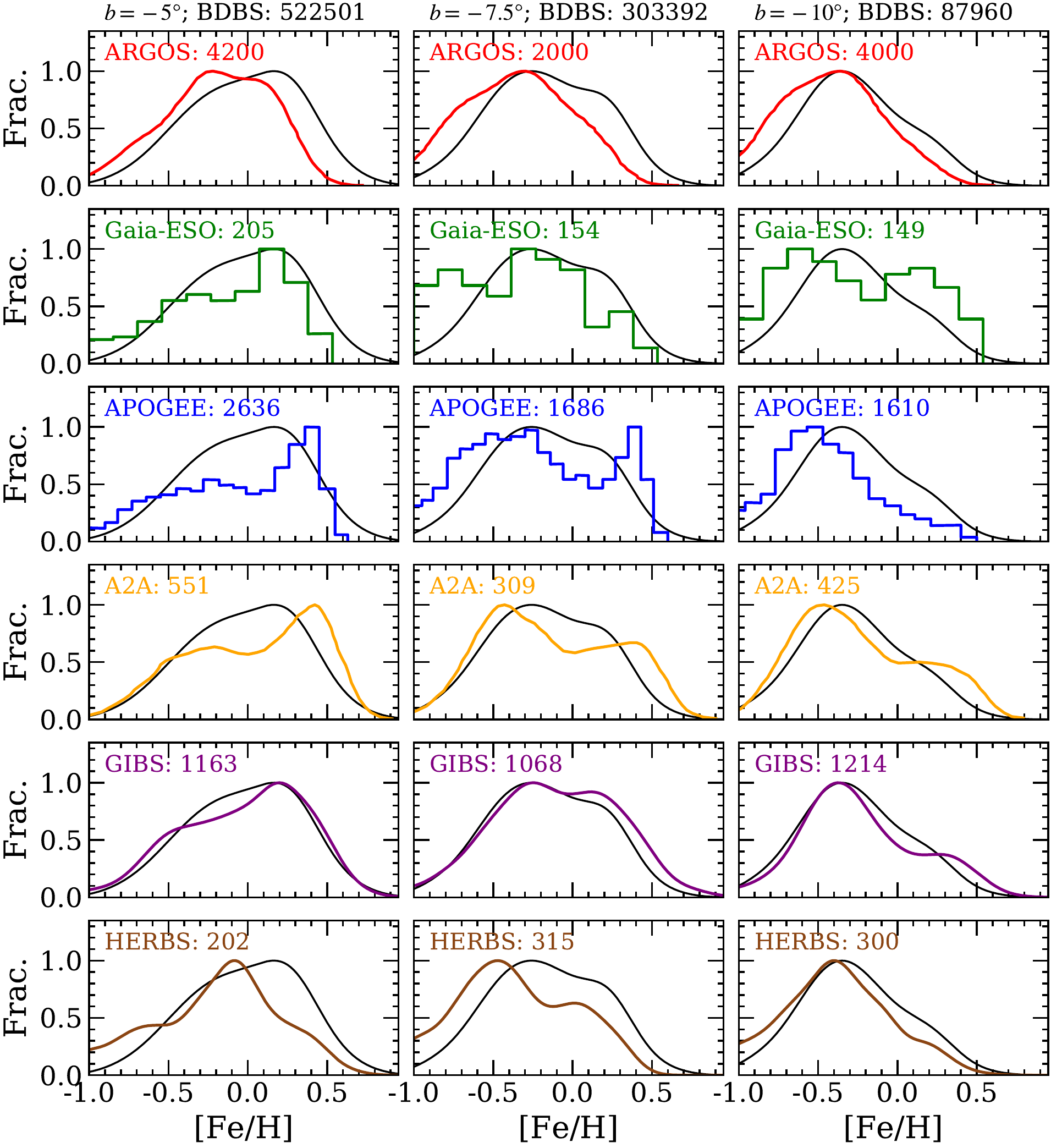}
\includegraphics[width=\columnwidth]{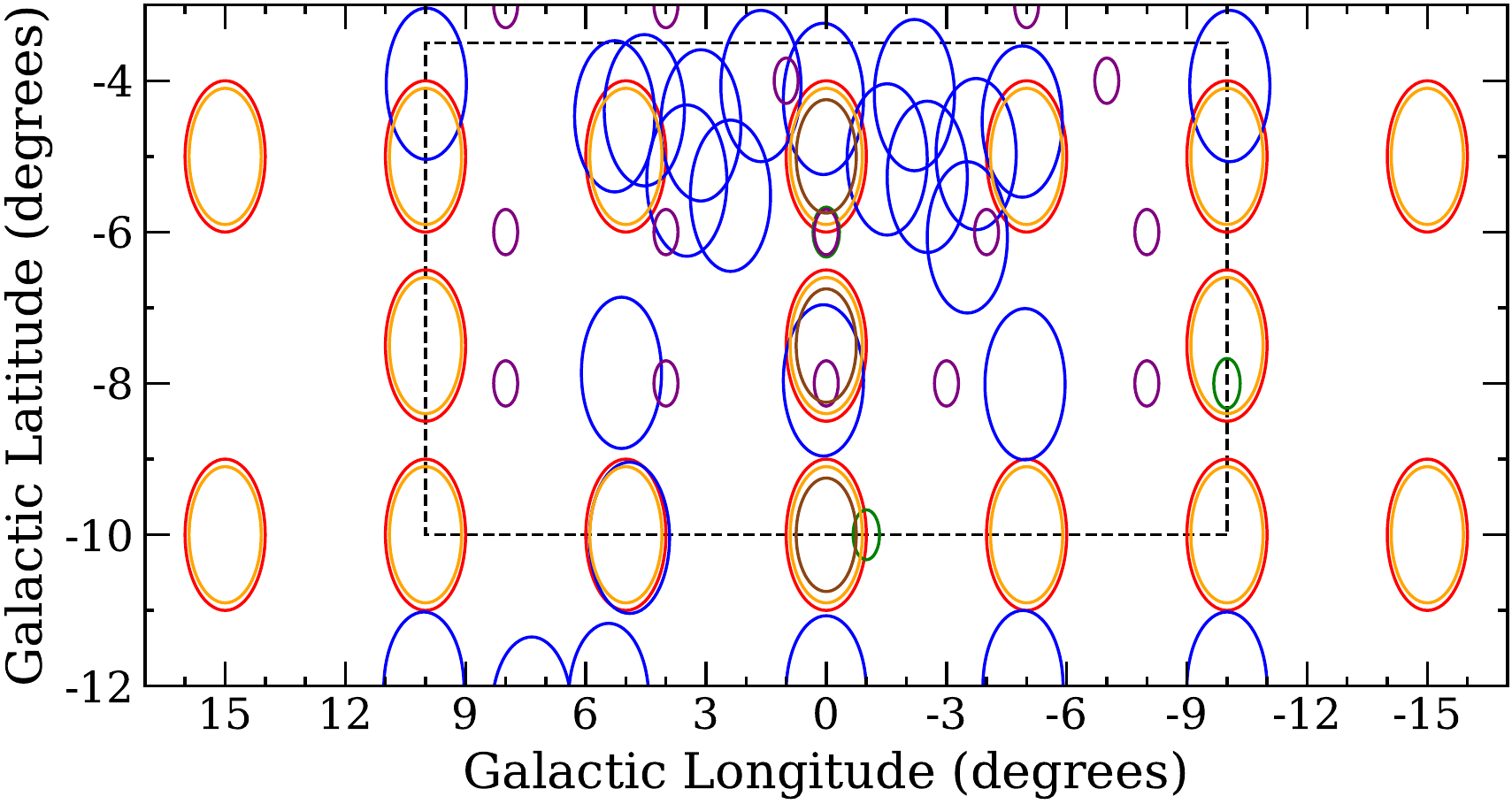}
\caption{\emph{Top panels:} BDBS metallicity distribution functions (black 
lines) are compared against similar data from the ARGOS 
\citep[red;][]{Ness13_mdf}, Gaia-ESO \citep[green;][]{Rojas14}, 
APOGEE \citep[blue;][]{Rojas20}, A2A \citep[orange;][]{Wylie21}, GIBS 
\citep[purple;][]{Zoccali17}, and HERBS \citep[brown;][]{Duong19} surveys.  
Histogram values for all data sets except GIBS and HERBS, which provided data 
tables, were measured off the published figures.  The number of stars in each 
distribution is provided with each panel.  The data cover all longitudes in
each sample when available, and generally span about $\pm$ 1$^{\circ}$ in 
latitude around $b = -5^{\circ}$ (left), $b = -7.5^{\circ}$ (middle), and 
$b = -10^{\circ}$ (right).  \emph{Bottom panel:} Approximate
field centers and sizes are outlined for each survey.  The
colors of the fields match the histograms in the top panels, and the dashed 
box highlights the approximate coverage of the BDBS data presented here.  For
overlapping sight lines, the field sizes have been slightly modified for 
visibility.}
\label{fig:mdf_comp}
\end{figure}

Figure \ref{fig:mdf_comp} compares the BDBS derived metallicity distribution 
functions against those in similar fields from the ARGOS, Gaia-ESO, APOGEE,
A2A, GIBS, and HERBS spectroscopic surveys.  Despite the variety in sample
sizes, data quality, and analysis methods, we can note several common patterns
that are reproduced by all surveys: (1) clear morphological changes in the 
metallicity distributions that are correlated with vertical distances from the 
Galactic plane; (2) most fields appear to have at least two local maxima; 
(3) all fields exhibit large metallicity ranges spanning 
at least $-1 <$ [Fe/H] $< +0.5$; and (4) the mean metallicity decreases with
increasing distance from the plane (i.e., a vertical metallicity gradient 
exists).  

Further examination of Figure \ref{fig:mdf_comp} reveals several important 
differences between the surveys.  For example, both the number of local maxima 
and their associated peak [Fe/H] locations are not consistent; however, at 
least some of this discrepancy is driven by differences in analysis
methods, data quality, and wavelengths.  These morphological differences are
particularly troublesome when comparing the ARGOS, A2A, and HERBS 
distributions, which have significant sample overlap.  Interestingly, the 
metallicity distributions appear to converge when observing fields farther 
from the plane.

For the inner bulge fields near $b = -5^{\circ}$, the metal-poor tails of all
seven surveys shown in Figure \ref{fig:mdf_comp} are surprisingly similar, 
at least for $-1 <$ [Fe/H] $< -0.4$.  Most of the surveys also show the same 
pattern of a dominant metal-rich peak near [Fe/H] $\sim$ $+$0.2.  However, 
the ARGOS and HERBS data sets have a paucity of very metal-rich stars.  The
Gaia-ESO data exhibit a truncated metal-rich tail compared to BDBS, APOGEE, and
GIBS while the A2A metal-rich peak appears shifted to a relatively high value
([Fe/H] $>$ $+$0.4).  We can also note that the global maximum for the HERBS
distribution occurs near the same metallicity ([Fe/H] $\sim$ $-$0.2) at which
the Gaia-ESO, APOGEE, and A2A data exhibit clear local minima.

At $b = -7.5^{\circ}$, the BDBS distribution is morphologically similar to 
those of the GIBS and A2A surveys.  Figure \ref{fig:mdf_comp} also indicates 
that the BDBS and HERBS distributions may be very similar when taking into 
account a possible $\sim$0.2 dex zero-point offset.  However, the ARGOS and 
Gaia-ESO data show significantly more stars with [Fe/H] $<$ $-0.8$.  Although
sample selection may explain part of this discrepancy, we note that most of
the surveys shown in Figure \ref{fig:mdf_comp} targeted red clump stars.
Similar to the $b = -5^{\circ}$ distributions, we note that the ARGOS sample 
has a much smaller number of very metal-rich stars compared to most other
surveys.  The metal-rich peak locations are also at least 0.1-0.2 dex higher
for the Gaia-ESO, APOGEE, and A2A distributions compared to those of BDBS, 
GIBS, and HERBS.

Unlike the fields closer to the plane, the $b = -10^{\circ}$ distributions
are all morphologically similar with strong metal-poor peaks near [Fe/H] $\sim$
$-0.4$ and long metal-rich tails.  The ARGOS, GIBS, A2A, and HERBS data sets
seem to require zero-point offsets of $\sim$0.1 dex or less to align with BDBS 
while those of Gaia-ESO and APOGEE are offset by $\sim$0.1-0.3 dex.  Although
the ARGOS metal-rich tail is now more similar to those of other surveys, 
the prominent very metal-poor peak at [Fe/H] $\sim$ $-0.8$ is not generally
observed in other data sets.  The Gaia-ESO distribution shows 
noticeably more stars with [Fe/H] $>$ 0 than any other survey, despite 
having a metal-rich peak at about the same location.

Finally, we can compare the mean metallicity gradient derived here against 
those in the literature.  As noted in Section \ref{sssec:gradients}, we find 
d[Fe/H]/d$b$ = $-0.05$ dex degree$^{-1}$ or d[Fe/H]/dZ = $-0.38$ dex 
kpc$^{-1}$ for fields with $-10^{\circ} < b < -3.5$ and  0.5 < |Z| < 1.5 kpc, 
respectively.  These values are in excellent agreement with previous estimates 
ranging from d[Fe/H]/d$b$ = $-0.04$ to $-0.075$ dex degree$^{-1}$ in observed 
coordinates \citep{Zoccali08,Gonzalez13,Rojas14} and spanning 
d[Fe/H]/dZ = $-0.6$ to $-0.24$ dex kpc$^{-1}$ in physical coordinates 
\citep{Zoccali08,Johnson13_offaxis,Ness13_mdf,Rojas17,Rojas20,Wylie21}.  
Figure \ref{fig:mdf_decomp} indicates that the metallicity gradient might 
become more shallow for |Z| $<$ 0.5 kpc, which would be in agreement with 
previous analyses \citep{Rich07_Inner,Rich12,Schultheis19,Rojas20}; however, 
our observations do not extend close enough to the plane to confirm this trend.

\subsection{The X-Shape in BDBS Data} \label{ssec:xshape}
\begin{figure*}
\includegraphics[scale=0.8]{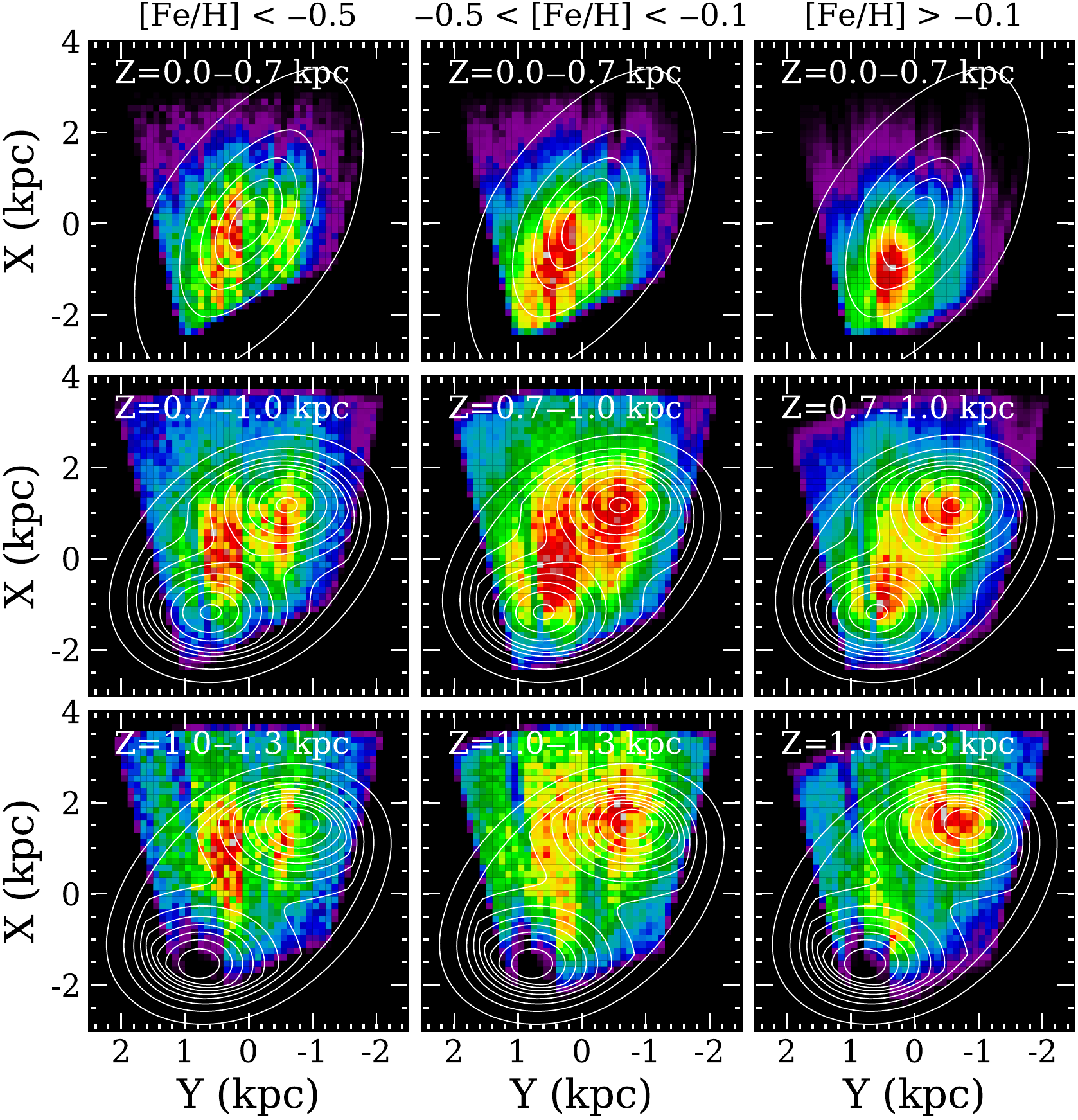}
\caption{The top, middle, and bottom rows show density map slices (as if viewing
the bulge from the North Galactic Pole) of BDBS red clump stars in physical
coordinate space with |Z| values ranging from 0.0-0.7 kpc, 0.7-1.0 kpc, and
1.0-1.3 kpc, respectively.  Similarly, the left, middle, and right columns only
include stars with [Fe/H] $<$ $-0.5$, $-0.5 <$ [Fe/H] $< -0.1$, and [Fe/H] $>$
$-0.1$, respectively.  In all panels, the highest density regions have red and
orange colors while the lowest density regions have black and purple colors.
For the top row, the contours show a slice of the best fit 3D density law that
best describes the VVV density distribution
\citep[][their model E, equation 2; see also their Table 3]{Simion17} projected
along the Z-axis between 0.0 and 0.7 kpc.  The contours for the middle and 
bottom rows show the projection of the X-shape density function provided by 
\citet{Lopez17}, which is a parameterization of the double red clump at 
intermediate latitudes based off of previous models by \citet{Lopez05}
and \citet{Wegg13}.}
\label{fig:feh_proj}
\end{figure*}

Using the observed Galactic coordinates in concert with the adopted 
relationships between [Fe/H], $M_{i,o}$, and distance outlined in 
Equations \ref{eq:abs_mag2}-\ref{eq:distance}, we can construct three 
dimensional density maps of the red clump distributions as functions of 
metallicity.  Figure \ref{fig:feh_proj} shows three vertical slices through
the density maps out to |Z| = 1.3 kpc and for three broad metallicity bins.
The figure also includes contour projections from \citet{Simion17} that fit
the VVV density distribution of the bulge/bar inside |Z| = 0.7 kpc and a 
separate parameterization of the X-shape structure from \citet{Lopez17} in the
outer bulge fields.

Visual inspection of Figure \ref{fig:feh_proj} shows that close to the plane
(|Z| $<$ 0.7 kpc) our observations are skewed to favor metal-rich ([Fe/H] $>$ 
$-0.5$) stars on the near side of the bar.  Although some previous studies 
found evidence of the X-shape structure emerging as low as $\sim$ 400 pc
from the plane \citep[e.g.,][]{Wegg13}, we did not find any X-shape signatures
for distances inside $\sim$ 700 pc.  The most metal-poor stars
in our sample ([Fe/H] $<$ $-0.5$) appear to be more uniformly distributed than
the more metal-rich stars, and exhibit a peak near the Galactic center (X = 0 
kpc).  This observation provides evidence that metal-poor stars in the 
bulge are distributed differently than those with [Fe/H] $>$ $-0.5$, as has 
been previously noted by several authors \citep[e.g.,][]{Ness13_mdf,Rojas14,
Zoccali17,Rojas20}.  However, we note that \citet{Kunder20} found evidence 
supporting two RR Lyrae populations co-existing in the bulge: one is 
constrained to within the inner 3.5 kpc of the Galactic Center and follows the 
bar while the other is more centrally concentrated but does not trace the bar. 
We suspect that a subset of our metal-poor sample traces the same population 
as the barred RR Lyrae, and that the fractional weights of the barred versus
non-barred groups vary with vertical distance from the plane.  Nevertheless, 
our observations reinforce previous velocity ellipsoid analyses that found a 
significant vertex deviation consistent with bar-like kinematics only exists 
for stars with [Fe/H] $>$ $-0.5$ \citep{Soto07,Babusiaux16,Simion21}.

At intermediate distances from the plane (0.7 $<$ |Z| $<$ 1.0 kpc), 
Figure \ref{fig:feh_proj} shows that the most metal-poor stars have 
little connection with the shape of the bar, and instead exhibit a more 
isotropic distribution centered near X = 0 kpc.  In contrast, the spatial
distributions for the intermediate and high metallicity bins strongly align 
with the modeled bar angle (20$^{\circ}$).  Stars with [Fe/H] $>$ $-0.5$
appear to cluster in two high density regions on the near and far sides of
the bar around X $\sim$ $\pm$ 1 kpc, similar to what \citet{Li12} found with
their N-body Milky Way bulge model.  In fact, the local maxima of these regions
are very similar in position and extent to the X-shape contours that are
evaluated between |Z| = 0.7-1.0 kpc.  The middle panel of 
Figure \ref{fig:feh_proj} provides evidence that the high density regions for 
stars with $-0.5 <$ [Fe/H] $< -0.1$ may be more extended along the 
Sun-Galactic Center (X) coordinate than higher metallicity stars.  However, 
this could also be a result of mixing stars that are more isotropically 
distributed with those of somewhat higher metallicity that are concentrated 
within the X-shape structure.  Alternatively, the larger extension in the X 
coordinate for intermediate metallicity stars could be a result of a thicker 
X-shape structure and/or ``contamination" from the far end of the X-shape.
We note also that \citet{Li15} showed that the X-shape is likely to 
have a broader, peanut-shaped morphology than a well-defined and thin
shape like the letter X.

For the largest vertical distance slice shown in Figure \ref{fig:feh_proj},
the metal-poor stars still appear more isotropically distributed, but the
highest density regions are found 0.5-1.0 kpc beyond the Galactic Center.
The shift in the peak position of metal-poor stars to larger distances is 
likely a result of the viewing angle along lines of sight that intersect
|Z| = 1.0-1.3 kpc, rather than an intrinsic change in the distribution.
Similar to the |Z| = 0.7-1.0 kpc vertical slices, the intermediate and 
metal-rich bins again show clear over-densities that are well-aligned with
both the bar angle and the X-shape model.  However, since
the BDBS map does not currently include measurements of stars with 
$b \la -9^{\circ}$ for fields with $l \ga +3^{\circ}$ \citep[e.g., see Figure
1 of ][]{Johnson20}, we are only able to partially detect the X-shape structure
on the near side of the bar \citep[see also][]{Gonzalez15_X}.   
The over-density on the far side of the bar almost exactly matches the model 
contours in location, extent, and shape.  Furthermore, the separation between 
the near and far side metal-rich over-densities becomes larger with increasing 
vertical distance, which suggests that we are directly observing the X-shape 
structure in the bulge.  


\subsection{Metallicity Dependent Spatial Distributions} \label{ssec:zdist}
\begin{figure*}
\includegraphics[width=\textwidth]{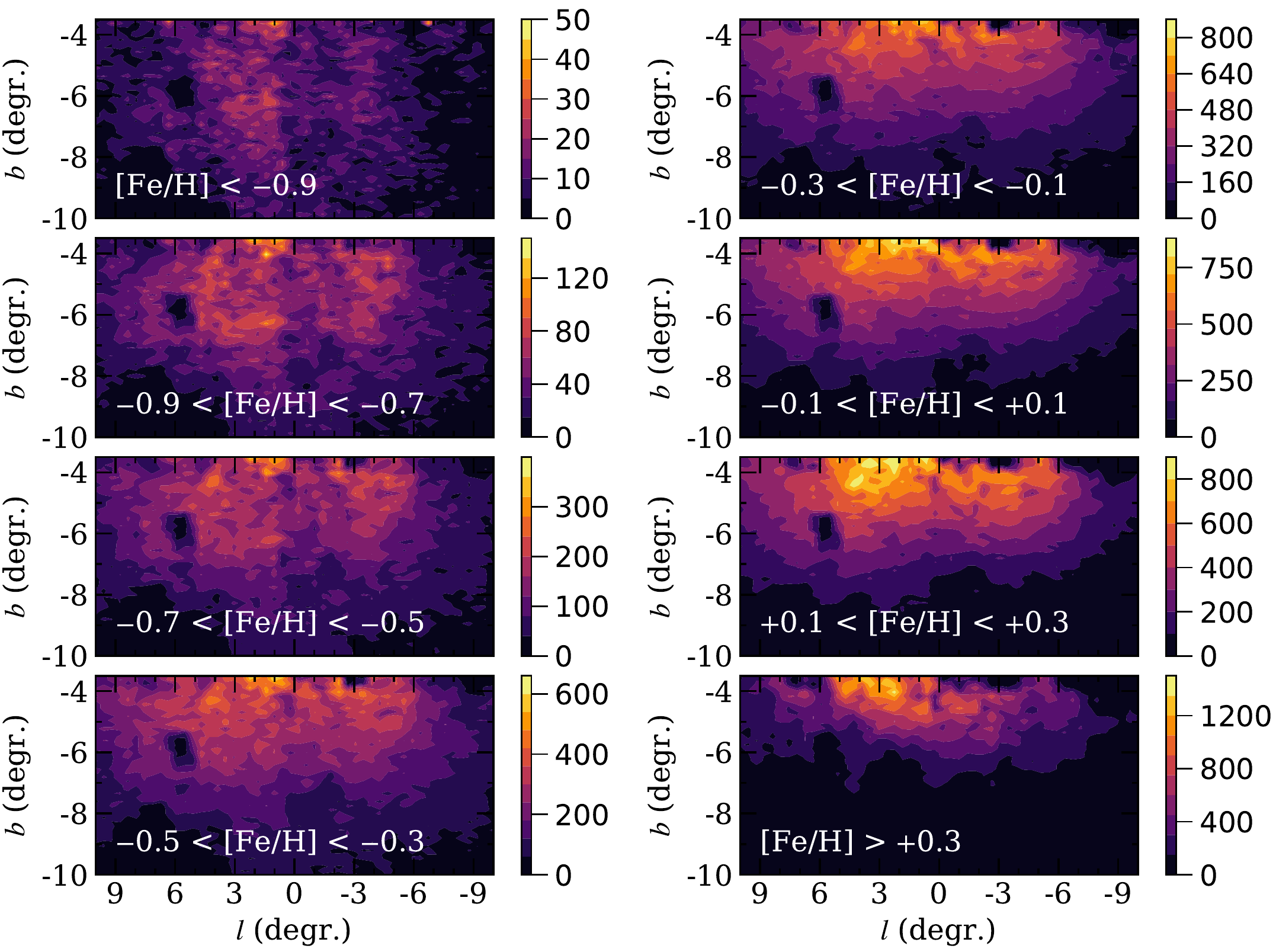}
\caption{Red clump density distribution contours are shown in observed
coordinates for several metallicity bins.  The total number of stars in each
panel ranges from $\sim$ 20,000 in the most metal-poor bin to $>$ 500,000 in
the most metal-rich bin.  A clear change in the spatial distribution is
observed for stars on either side of [Fe/H] $\sim$ $-0.3$.  The metal-poor
stars extend to high latitudes and appear to exhibit a more spheroidal
distribution.  In contrast, the metal-rich stars are concentrated
close to the plane and exhibit a boxy/peanut morphology.}
\label{fig:feh_lb}
\end{figure*}

Previous surveys have established that the observed metallicity gradient is
likely driven by changes in the vertical scale height of bulge stars with
different mean metallicities \citep[e.g.,][]{Zoccali08,Hill11,Ness13_mdf,
Zoccali17,Duong19,Rojas20}.  In Figure \ref{fig:feh_lb}, we show the spatial
distribution of red clump stars in the BDBS survey area with different
metallicities, and confirm that the vertical scale heights change strongly
as a function of metallicity.

Further visual inspection of Figure \ref{fig:feh_lb} reveals strong 
morphological changes in the spatial distributions as a function of 
metallicity as well.  For example, stars with [Fe/H] $<$ $-0.3$ almost uniformly
fill the BDBS footprint area and lack any clear structure.  However, we 
systematically observe about 5 per cent more stars with [Fe/H] $<$ $-0.3$ at 
positive longitudes than negative longitudes, which suggests that the 
metal-poor stars may form an oblate spheroid that is at least weakly aligned
with the bar.  Starting with the $-0.3 <$ [Fe/H] $< -0.1$ bin in 
Figure \ref{fig:feh_lb}, the outer contours become more rounded and/or 
peanut-shaped while simultaneously exhibiting a retreat to smaller Galactic 
latitude values.  The peanut shape is less obvious in the highest metallicity
bin, but this is likely due to the strong concentration near the plane for these
stars coupled with our limited observations inside $b = -4^{\circ}$.  
Additionally, these stars may form a more ``central boxy core" that is less
peanut-shaped, as discussed in \citet{Li15}.  The metallicity dependent
spatial variations observed here are consistent with previous investigations
of Mira stars that found the older, more metal-poor stars traced a spheroid 
while the younger, more metal-rich population exhibited a barred structure
\citep{Catchpole16,Grady20}.

\begin{figure*}
\includegraphics[width=1.0\textwidth]{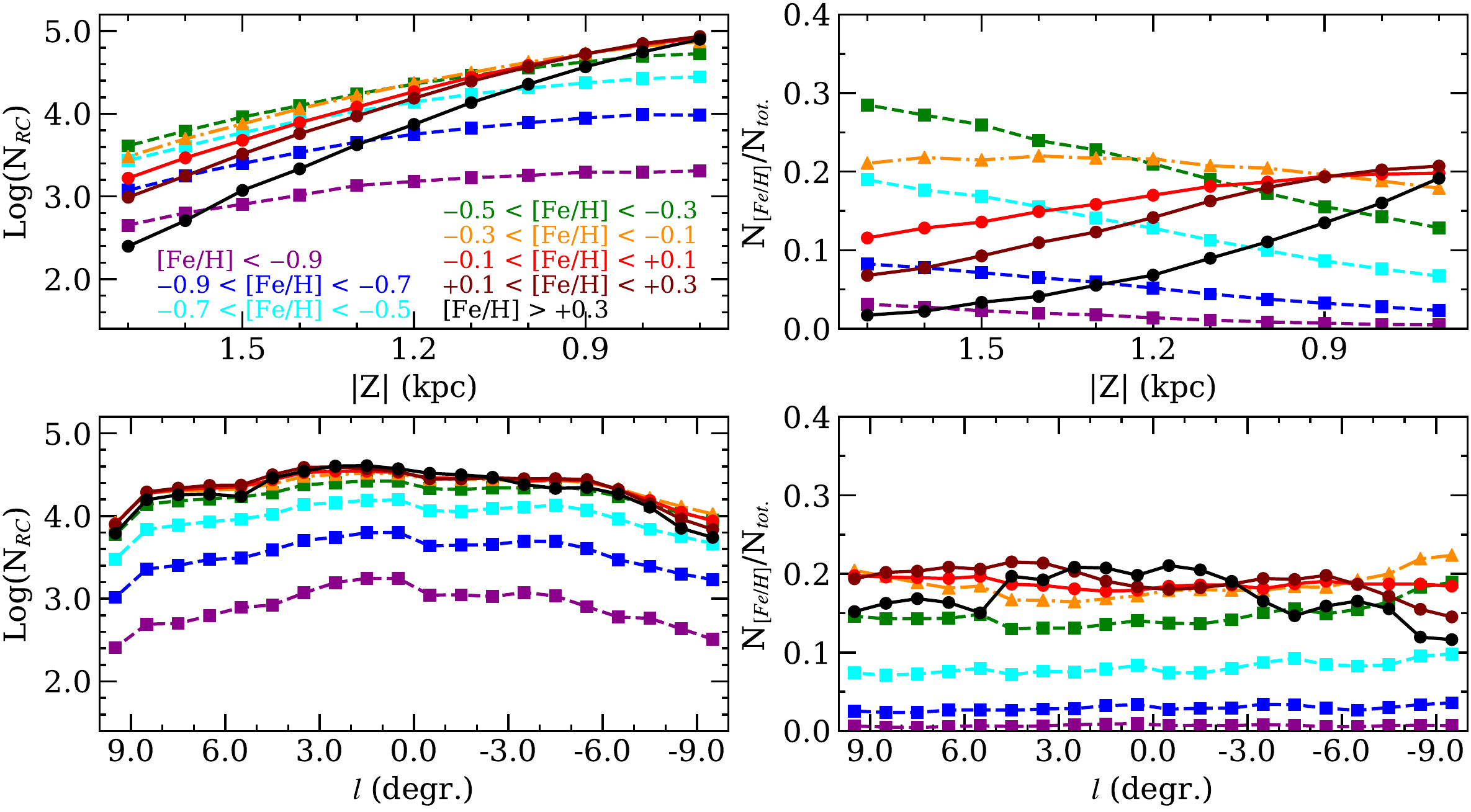}
\caption{The top left panel plots the log of the number of red clump stars
in 0.1 kpc vertical stripes, summed across all longitudes, for the 
metallicity bins shown in Figure \ref{fig:feh_lb}.  The bottom left panel
plots the ratio of the number of red clump stars within a given metallicity 
bin against the total number of red clump stars in each vertical stripe.
This panel highlights two broad ``families" that show different dependencies 
with distance from the plane, which are also correlated with [Fe/H].  Groups
with decreasing fractions at lower latitudes are indicated with filled boxes
and dashed lines, those with a flat functional form are indicated with filled
triangles and dot-dashed lines, and those with increasing functions have filled
circles and solid lines.  The bottom panels show similar plots as a function of
Galactic longitude, summed across all latitudes, and highlight that no 
similar metallicity-dependent gradient is found in the radial direction.}
\label{fig:feh_scale_height2}
\end{figure*}

Changes to the vertical distributions of bulge stars are further illustrated 
in Figure \ref{fig:feh_scale_height2}, which plots the 
red clump star counts and fractional compositions as functions of 
vertical distance from the plane.  We find that the number densities of stars 
with [Fe/H] $<$ $-0.3$ only change by about a factor of 2-10 between the most 
inner and outer fields sampled in Figure \ref{fig:feh_scale_height2}.  This 
observation is consistent with previous assertions that most of the metal-poor 
stars in the bulge likely form  either an oblate spheroid or ``thick bar" 
distribution, which is also supported by kinematic 
arguments \citep[e.g.,][]{Babusiaux10}.  A combination of the observed 
spatial distribution and shifting metal-poor peak position with latitude
may be useful for constraining dissipative collapse or gas outflow parameters
in future models.

In contrast to the metal-poor vertical distributions, 
Figure \ref{fig:feh_scale_height2} shows that more metal-rich stars exhibit 
substantial increases in red clump number densities when moving from the 
outer to inner bulge, and that the rate of increase is strongly correlated
with chemical composition.  For example, the number of red clump stars with 
$-0.1 <$ [Fe/H] $< +0.1$ increases by about a factor of 60 between |Z| = 1.7 and 
0.7 kpc.  However, over the same distance the number of red clump stars with 
[Fe/H] $>$ $+0.3$ increases by a factor of 400.

The strong correlation between vertical slice star counts and metallicity is 
also reflected in the right panels of Figure \ref{fig:mdf_decomp}, which show 
that the metal-rich tails becomes broader closer to the plane.  Such 
observations may reflect variations in either the star formation efficiency or 
gas infall times in the bulge, which are likely to increase the width of the 
resulting metallicity distribution functions without shifting the metal-rich 
peak position \citep[e.g.,][]{Ballero07}.

The top right panel of Figure \ref{fig:feh_scale_height2}, which shows the 
change in the ratios of stars within a given metallicity bin to the total 
number of stars in a vertical slice, also support the presence of two broad 
stellar groups.  The metallicity bins that include stars with 
[Fe/H] $<$ $-0.3$ all exhibit the same basic pattern where their fractional 
contributions have maximums in the outer bulge and then monotonically 
decrease when moving closer to the plane.  Similarly, all of the metallicity 
bins that include stars with [Fe/H] $> -0.1$ increase their 
$N_{[Fe/H]}/N_{tot.}$ ratios monotonically from the outer to inner bulge.  

The $-0.3 <$ [Fe/H] $< -0.1$ metallicity bin maintains a nearly constant
$N_{[Fe/H]}/N_{tot.}$ ratio of about 0.2, and signals a transition between 
the two groups.  This metallicity bin aligns roughly with the [$\alpha$/Fe]
inflection point found near [Fe/H] $\sim$ $-0.3$ \citep[e.g.,][]{Fulbright07,
AlvesBrito10,Gonzalez11,Johnson11,Bensby17,Rojas17,Duong19}, and also with the 
vertex deviation at [Fe/H] $\sim$ $-0.5$ observed in some minor-axis bulge 
fields \citep{Zhao94,Soto07,Babusiaux10,Soto12,Simion21}.  The bottom
panels of Figure \ref{fig:feh_scale_height2} highlight the lack of any
similar metallicity gradient in the radial (longitudinal) direction.

A comparison of the metal-poor and metal-rich red clump 
distributions shown in Figure \ref{fig:feh_scale_height2} with the Gaussian 
decompositions presented in \citet{Zoccali17} leads to an apparent conflict of 
results.  Extrapolating the star counts of each metallicity bin in 
Figure \ref{fig:feh_scale_height2} from the last reference point toward the 
plane should result in a continued increase in the prominence of stars with
[Fe/H] $>$ $-0.1$ concurrent with a strong (relative) decrease in the fraction 
of stars with [Fe/H] $<$ $-0.5$.  However, Figure 7 of \citet{Zoccali17} shows  
that the metal-poor component decreases in fractional contribution between 
$b = -8.5^{\circ}$ and $-3.5^{\circ}$, and then increases again at latitudes 
closer to the plane.

Although the BDBS data cannot rule out an increase in the metal-poor star 
count trends for fields with $b > -3.5^{\circ}$, a possible explanation, as
mentioned in Section \ref{sssec:mdf_models}, is that the metal-poor and 
metal-rich groups are not well-represented as Gaussian distributions.  
Instead, these broad groups are likely skewed distributions with long tails, 
which can lead to over/under-counting fractional compositions when using 
Gaussian mixture models.  If we assume the |Z| = 0.5 and 2.0 kpc distributions 
of Figure \ref{fig:mdf_bdz} are close representations of the true metal-poor and
metal-rich groups, respectively, then we find that sight lines with strong
metal-rich peaks will naturally contain substantial numbers of metal-poor 
stars.  Furthermore, as mentioned in Section \ref{sssec:tail_variations}, we 
also find evidence that the metal-rich tail becomes broader closer to 
the plane, and especially along the minor-axis.  A broadening of the 
metal-rich metallicity distribution function might also strengthen the 
population's metal-poor tail such that it mimics a growing contribution from 
the separate metal-poor group.

Therefore, we posit that the large population of metal-poor stars found in 
fields close to the plane by \citet{Zoccali17} may be driven more by the 
metal-rich group's metal-poor tail than the rising prominence of a truly 
separate metal-poor population.  In fact, we argue that the bulge may contain
two distinct metal-poor populations: one group is the metal-poor tail of 
the star formation event that produced the metal-rich peak while the other
is the vertically extended spheroidal population that dominates at high 
latitudes.  This assertion is consistent with the chemodynamical results
presented by \citet{Queiroz20} that indicate two or more metal-poor 
populations co-existing in the bulge that possess different kinematic
signatures.

We also note \citet{Zoccali17} found that while metal-poor stars have higher 
velocity dispersions than metal-rich stars in the outer bulge, the two 
populations have similar velocity dispersions between $b = -3.5^{\circ}$ and 
$-2.0^{\circ}$.  \citet{Zoccali17} attributed this result to the metal-poor 
population hosting a velocity gradient; however, one might also expect a 
similar observation if a majority of stars within a few degrees of the plane 
are actually part of the same population that produced the metal-rich 
peak\footnote{Note that \citet{Zoccali17} also find for a single pointing of 
a minor-axis $b=-1^{\circ}$ field that the metal-poor velocity dispersion continues to 
decrease below the level observed for metal-rich stars.}.  If the inner few 
degrees of the bulge are dominated by a single long-tailed distribution then 
this would support previous observations showing that the metallicity gradient 
flattens out and does not extend to the inner bulge \citep{Rich07_Inner,
Rich12,Schultheis19,Rojas20}. 

\section{Summary} \label{sec:summary}
For this work, we have isolated a sample of $\sim$ 2.6 million red clump stars
spanning $|l| < 10^{\circ}$ and $-10^{\circ} < b < -3.5^{\circ}$ from the 
BDBS catalog \citep{Johnson20,Rich20}.  We applied Gaia EDR3 proper
motions and parallaxes to remove likely foreground contamination, and also 
determined distance estimates from a metallicity-dependent calibration of 
dereddened $i$-band photometry based on 10 Gyr stellar isochrones.  
Metallicities were determined for each red clump star using the calibration 
between BDBS $(u-i)_{o}$ colors and spectroscopic GIBS [Fe/H] values presented 
in \citet{Johnson20}.  Using these data, we derived metallicity distribution
functions and three dimensional density maps over most of the BDBS footprint.
A summary of the results includes:

\begin{itemize}
\item In agreement with past work, we find a strong change in the metallicity
distribution function morphology that varies with vertical distance from
the plane.  However, the metallicity distribution functions generally have a 
weak dependence on longitude.

\item Metallicity distributions for fields close to the plane ($b \sim 
-4^{\circ}$) are asymmetric, strongly peaked near [Fe/H]~$\sim$~$+$0.2, and 
possess a long metal-poor tail.  A clear second and more metal-poor peak
emerges starting at $b \sim -5^{\circ}$ and grows in prominence with increasing
distance from the plane.  Additionally, while the metal-rich peak remains 
fixed near [Fe/H] = $+$0.2 for all latitudes, the metal-poor peak shifts from 
[Fe/H] $\sim$ $-0.15$ to $-0.4$ between $b \sim -4^{\circ}$ and $-10^{\circ}$
(0.5 $<$ |Z| 1.5 kpc), respectively.  A third, more metal-poor peak near 
[Fe/H] $\sim$ $-0.8$ is found along a few sight lines but only for |Z| $\sim$
2 kpc.

\item Overall, the mean metallicity decreases from [Fe/H] = 0 at 
$b = -3.75^{\circ}$ to [Fe/H] = $-0.3$ at $b = -9.75^{\circ}$.  Similar results
are found if the data are binned by physical distance from the plane (Z) rather
than observed Galactic latitude.  The main difference when binning by Z is 
that these metallicity distribution functions appear to have somewhat sharper
peaks and tails, likely due to the removal of stars at different Z distances
that ``contaminate" the distribution when integrating along a line of constant 
Galactic latitude.

\item Using the inner and outermost fields as representations of the true
underlying metal-poor and metal-rich populations, we find that the metallicity
distribution function morphologies are likely not well-represented by Gaussian
functions.  Instead, the two components may be more similar to the asymmetric,
long-tailed morphologies generated either by chemical enrichment models or 
from empirical local thin and thick disk metallicity distribution functions.
However, the bulge populations are generally at least 0.2 dex more metal-rich
than the local thin and thick disk stars.

\item Similar to previous studies, we find a clear vertical metallicity
gradient equivalent to $-0.05$ dex degree$^{-1}$ ($-0.38$ dex kpc$^{-1}$) for 
sight lines between $b = -3.5^{\circ}$ and $-10^{\circ}$ (|Z| = 0.5 to 1.5 
kpc).  Additionally, we find a clear negative gradient for the metal-poor peak
position of $-0.04$ dex degree$^{-1}$ ($-0.37$ dex kpc$^{-1}$), which is 
nearly identical to the mean metallicity gradient.  The observed correlation
between the metal-poor peak position and distance from the plane could have
been driven by galaxy-scale mechanisms, such as winds/outflows or inside-out
star formation, rather than accretion or the tidal destruction of globular 
clusters.

\item An interesting new result presented here is the correlation
between the metal-rich tail width and distance from the plane.  We find that
the metal-rich tails truncate more sharply after the peak in higher latitude
fields compared to those closer to the plane.  Additionally, we find 
in fields with $b > -6^{\circ}$ that sight lines along the minor axis 
($|l| < 3^{\circ}$) have longer metal-rich tails than those
farther from the Galactic Center.  

\item Extracting slices of constant Z through the BDBS data set revealed
clear evidence of metallicity dependent variations in the spatial density
distributions.  In agreement with past work, stars with [Fe/H] $<$ $-0.5$ show 
only a weak connection to the bar and appear more isotropically distributed 
than their more metal-rich counterparts.  The stars with [Fe/H] $<$ $-0.5$ 
also do not show any evidence of participating in orbits that support the 
X-shape structure.  In contrast, stars with [Fe/H] $>$ $-0.5$ follow the bar
orientation at all vertical distances.

\item For fields with |Z| $>$ 0.7 kpc, stars with [Fe/H] $>$ $-0.5$ exhibit
two spatially concentrated, high density regions on either side of the Galactic 
Center that align with the bar angle.  Furthermore, the separation between the
two over-densities increases with distance from the plane, and a comparison 
between the observed distributions and a parametric model reveals that these 
over-densities match the predicted contours of the X-shape structure.  

\item Metallicity dependent red clump density contour maps reveal a strong 
change in the 2D spatial distribution of stars as a function of metallicity
as well.  Stars with [Fe/H] $<$ $-0.3$ span the entire BDBS footprint and 
appear to form an oblate spheroid or thick bar distribution.  In contrast,
starting at [Fe/H] $\sim$ $-0.3$ stars begin to form a clear boxy/peanut
shape structure that is associated with the inner disk/bar.  However, the 
peanut shape is less defined for stars with [Fe/H] $>$ $+$0.3 that are 
close to the plane, which may suggest that the inner portion of the bulge
forms more of a central boxy core, as noted in \citet{Li15}.  We also find
that the vertical extent of stars rapidly decreases with increasing 
metallicity, especially for [Fe/H] $>$ 0.

\item In general, we find evidence of two broad distributions where stars with 
[Fe/H] $<$ $-0.3$ exhibit declining fractional contributions
when moving closer to the plane while those with [Fe/H] $>$ $-0.1$ exhibit
increasing fractional contributions.  Stars with $-0.3 <$ [Fe/H] $< -0.1$ serve
as a transition point and exhibit a nearly constant N$_{[Fe/H]}/N_{tot.}$ 
ratio of about 0.2 across all latitudes investigated here.

\item The vertical number density distributions are also sensitive functions of
[Fe/H].  For example, the number of stars with [Fe/H] $<$ $-0.3$ only 
increases by about a factor of 2-10 between $b = -10^{\circ}$ and $-4^{\circ}$.
However, stars with $-0.1 <$ [Fe/H] $< +0.1$ increase by a factor 60 and stars 
with [Fe/H] $>$ $+$0.3 increase by a factor of 400 over that same vertical 
range.

\item The monotonically decreasing fractional contributions of metal-poor 
stars with decreasing distance from the plane seems to contrast with the 
results of \citet{Zoccali17}, which showed that the metal-poor population
reaches a minimum near $|b|$ $\sim$ 3.5$^{\circ}$ and increases when moving
closer to the plane.  We argue instead that the bulge has two metal-poor
populations: one that forms the long metal-poor tail of the distribution
that peaks near [Fe/H] = $+$0.2 and the other that belongs to the metal-poor
population that dominates in the outer bulge.  The existence of two or more
metal-poor groups is supported by chemodynamical data presented in 
\citet{Queiroz20}.  We note also that the inner bulge being mostly dominated
by a single asymmetric, long-tailed metallicity distribution would help explain 
observations finding a weak or null metallicity gradient within a few degrees 
of the plane.
\end{itemize}

The photometric metallicity maps presented here offer new insight into the 
bulge's formation history, and in the future may be combined with 
future Gaia data releases to obtain accurate chemodynamical maps encompassing
millions of red clump stars.  These data also offer a glimpse at the type of
science that may be achieved with upcoming observations from the Vera C. Rubin
Observatory, and highlight the critical role wide-field $u$-band observations
can fill for reconstructing the Galaxy's formation history.

\clearpage
\section*{Acknowledgements}
The authors gratefully acknowledge the anonymous referee for a careful
reading of the manuscript and for providing constructive criticism that
improved the work.
R.M.R. acknowledges his late father Jay Baum Rich for financial support.
C.A.P. acknowledges the generosity of the Kirkwood 
Research Fund at Indiana University.  A.M.K. acknowledges support from grant 
AST-2009836 from the National Science Foundation.  A.J.K.H. gratefully 
acknowledges funding 
by the Deutsche Forschungsgemeinschaft (DFG, German Research Foundation) -- 
Project-ID 138713538 -- SFB 881 (``The Milky Way System''), subprojects A03, 
A05, A11.  The research presented here is partially supported by the National 
Key R\&D Program of China under grant No. 2018YFA0404501; by the National 
Natural Science Foundation of China under grant Nos. 12025302, 11773052, 
11761131016; by the ``111'' Project of the Ministry of Education of China 
under grant No. B20019; and by the Chinese Space Station Telescope project.
This research was supported in part by Lilly Endowment, Inc., 
through its support for the Indiana
University Pervasive Technology Institute, and in part by the Indiana METACyt 
Initiative. The Indiana METACyt Initiative at IU was also supported in part by 
Lilly Endowment, Inc.  This material is based upon work supported by the 
National Science Foundation under Grant No. CNS-0521433.  This work was 
supported in part by Shared University Research grants from IBM, Inc., to 
Indiana University.  This project used data obtained with the Dark Energy 
Camera (DECam), which was constructed by the Dark Energy Survey (DES) 
collaboration.  Funding for the DES Projects has been provided by the U.S. 
Department of Energy, the U.S. National Science Foundation, 
the Ministry of Science and Education of Spain, the Science and Technology 
Facilities Council of the United Kingdom, the Higher Education Funding Council 
for England, the National Center for Supercomputing Applications at the 
University of Illinois at Urbana-Champaign, the Kavli Institute of Cosmological
Physics at the University of Chicago, the Center for Cosmology and 
Astro-Particle Physics at the Ohio State University, the Mitchell Institute for
Fundamental Physics and Astronomy at Texas A\&M University, Financiadora de 
Estudos e Projetos, Funda{\c c}{\~a}o Carlos Chagas Filho de Amparo {\`a} 
Pesquisa do Estado do Rio de Janeiro, Conselho Nacional de Desenvolvimento 
Cient{\'i}fico e Tecnol{\'o}gico and the Minist{\'e}rio da Ci{\^e}ncia, 
Tecnologia e Inovac{\~a}o, the Deutsche Forschungsgemeinschaft, 
and the Collaborating Institutions in the Dark Energy Survey.  The 
Collaborating Institutions are Argonne National Laboratory, 
the University of California at Santa Cruz, the University of Cambridge, 
Centro de Investigaciones En{\'e}rgeticas, Medioambientales y 
Tecnol{\'o}gicas-Madrid, the University of Chicago, University College London, 
the DES-Brazil Consortium, the University of Edinburgh, 
the Eidgen{\"o}ssische Technische Hoch\-schule (ETH) Z{\"u}rich, 
Fermi National Accelerator Laboratory, the University of Illinois at 
Urbana-Champaign, the Institut de Ci{\`e}ncies de l'Espai (IEEC/CSIC), 
the Institut de F{\'i}sica d'Altes Energies, Lawrence Berkeley National 
Laboratory, the Ludwig-Maximilians Universit{\"a}t M{\"u}nchen and the 
associated Excellence Cluster Universe, the University of Michigan, 
{the} National Optical Astronomy Observatory, the University of Nottingham, 
the Ohio State University, 
the OzDES Membership Consortium
the University of Pennsylvania, 
the University of Portsmouth, 
SLAC National Accelerator Laboratory, 
Stanford University, 
the University of Sussex, 
and Texas A\&M University.  Based on observations at Cerro Tololo 
Inter-American Observatory, National Optical Astronomy Observatory 
(2013A-0529;2014A-0480; R.M. Rich), which is operated by the Association of 
Universities for Research in Astronomy (AURA) under a cooperative agreement 
with the National Science Foundation.  This work has made use of data from the 
European Space Agency (ESA) mission {\it Gaia} 
(\url{https://www.cosmos.esa.int/gaia}), processed by the {\it Gaia}
Data Processing and Analysis Consortium (DPAC,
\url{https://www.cosmos.esa.int/web/gaia/dpac/consortium}). Funding for the 
DPAC has been provided by national institutions, in particular the institutions
participating in the {\it Gaia} Multilateral Agreement.

\section*{Data Availability}
\noindent
The raw and pipeline reduced DECam images are available for 
download on the NOAO archive at http://archive1.dm.noao.edu/.  Astrometric,
photometric, and reddening catalogs are in the process of being prepared for
public release.  The data used in this paper are provided in an electronic 
table, but further information, including an extended release of BDBS data
beyond what is presented here, may be provided upon request to the 
corresponding author.




\bibliographystyle{mnras}
\bibliography{references}


\clearpage
\begin{table*}
\tiny
\caption{Bulge Red Clump Data}
\label{tab:data_table}
\begin{tabular}{ccccccccccccccc}
\hline
RA      &       DEC     & $u$	&	$u_{err}$       &       $g$ &	$g_{err}$	&       $i$ &	$i_{err}$	&       $A_{u}$	& $A_{g}$	&	$A_{i}$	& Dist.  &       $\sigma$ Dist.	&	[Fe/H]	&	$\sigma$ [Fe/H]	\\
(degrees)       &       (degrees)       &       (mag.)  &       (mag.)  &       (mag.)  &       (mag.)  &       (mag.)  &       (mag.)  & (mag.)  &       (mag.)  &       (mag.)  & (kpc)   &	(kpc)	&       (dex)	& (dex)   \\
\hline
263.73815 & $-$39.34179 & 22.866 & 0.229 & 19.830 & 0.007 & 17.064 & 0.016 & 4.682 & 3.738 & 2.030 & 8.731 & 0.858 & $-$0.30 & 0.49 \\
263.81260 & $-$39.36577 & 22.640 & 0.037 & 19.964 & 0.003 & 17.220 & 0.007 & 5.204 & 4.154 & 2.256 & 8.754 & 0.886 & $-$0.68 & 0.53 \\
263.83048 & $-$39.34130 & 22.891 & 0.451 & 20.358 & 0.000 & 17.639 & 0.006 & 5.073 & 4.050 & 2.200 & 10.951 & 1.150 & $-$0.73 & 0.58 \\
263.83091 & $-$39.36591 & 22.933 & 0.235 & 20.543 & 0.012 & 17.732 & 0.006 & 5.236 & 4.180 & 2.270 & 11.146 & 1.147 & $-$0.82 & 0.55 \\
263.83457 & $-$39.34498 & 23.077 & 0.246 & 20.173 & 0.002 & 17.342 & 0.006 & 5.041 & 4.024 & 2.186 & 9.366 & 0.949 & $-$0.45 & 0.53 \\
... &       ...     &       ... &       ...   &       ... &       ...   &       ...  &       ...   &       ...   &       ...   &       ...   &       ... &       ... & ... &       ...    \\
280.70911 & $-$24.54837 & 19.553 & 0.001 & 17.440 & 0.008 & 16.090 & 0.003 & 1.146 & 0.915 & 0.497 & 11.492 & 0.769 & $-$0.49 & 0.12 \\
280.71047 & $-$24.54482 & 18.874 & 0.004 & 16.538 & 0.002 & 15.127 & 0.017 & 1.146 & 0.915 & 0.497 & 7.269 & 0.490 & $-$0.33 & 0.12 \\
280.71134 & $-$24.54170 & 19.413 & 0.001 & 17.173 & 0.001 & 15.821 & 0.025 & 1.146 & 0.915 & 0.497 & 10.086 & 0.686 & $-$0.42 & 0.12 \\
280.71366 & $-$24.54457 & 19.690 & 0.017 & 17.066 & 0.008 & 15.664 & 0.001 & 1.146 & 0.915 & 0.497 & 9.175 & 0.614 & $-$0.17 & 0.12 \\
280.71960 & $-$24.51606 & 19.308 & 0.016 & 17.261 & 0.017 & 15.933 & 0.029 & 1.187 & 0.948 & 0.515 & 10.665 & 0.732 & $-$0.55 & 0.12 \\
\hline
\multicolumn{15}{c}{The full version of this table is provided in electronic form.}
\end{tabular}
\label{tab:rc_data}
\end{table*}

\appendix

\section{Luminosity and Color Cuts for Red Clump Stars} \label{sec:appendix}

\begin{figure*}
\includegraphics[width=\textwidth]{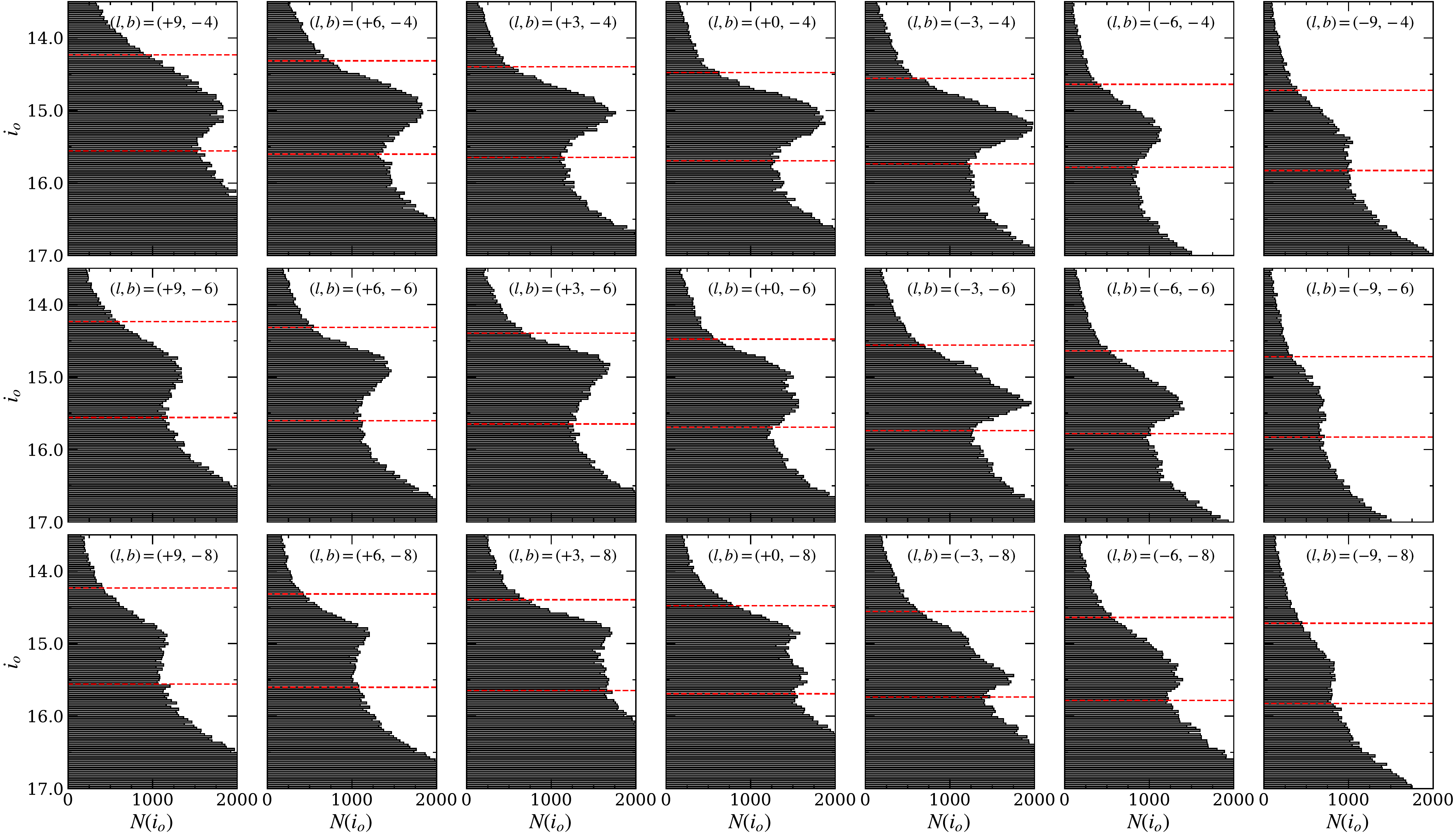}
\caption{Similar to the middle panels of Figure \ref{fig:cmd_selection}, 
$i_{o}$ magnitude selection limits (dashed red lines) for bulge red clump 
stars are shown for a variety of fields spanning a large fraction of the 
BDBS footprint.  The selection limits are calculated from 
Equations \ref{eq:bright_lim} and \ref{eq:faint_lim}.  The panels indicate 
that the selection criteria encapsulate the red clump region, regardless of 
the sight line, by taking into account the X-shape (double red clump) and bar 
angle.  The field radii range from 30$\arcmin$ in the $b = -4^{\circ}$ fields 
to 1$^{\circ}$ in the $b = -8^{\circ}$ fields.}
\label{fig:app_cmd_selection_grid}
\end{figure*}

We provide additional examples of luminosity and color functions in the red 
clump regions of 21 fields spanning a large fraction of the BDBS footprint.  
The dashed red lines in Figure \ref{fig:app_cmd_selection_grid} show the 
selection criteria when adopting the bright and faint magnitude limits 
presented in Equations \ref{eq:bright_lim} and \ref{eq:faint_lim}.  The data
show that we sufficiently encapsulate the red clump region, regardless of 
whether a single or double red clump feature is present.  Figure 
\ref{fig:app_cmd_selection_grid} also shows the expected behavior of the 
red clump becoming fainter at more negative longitudes, which is consistent
with the bar viewing angle.

\begin{figure*}
\includegraphics[width=\textwidth]{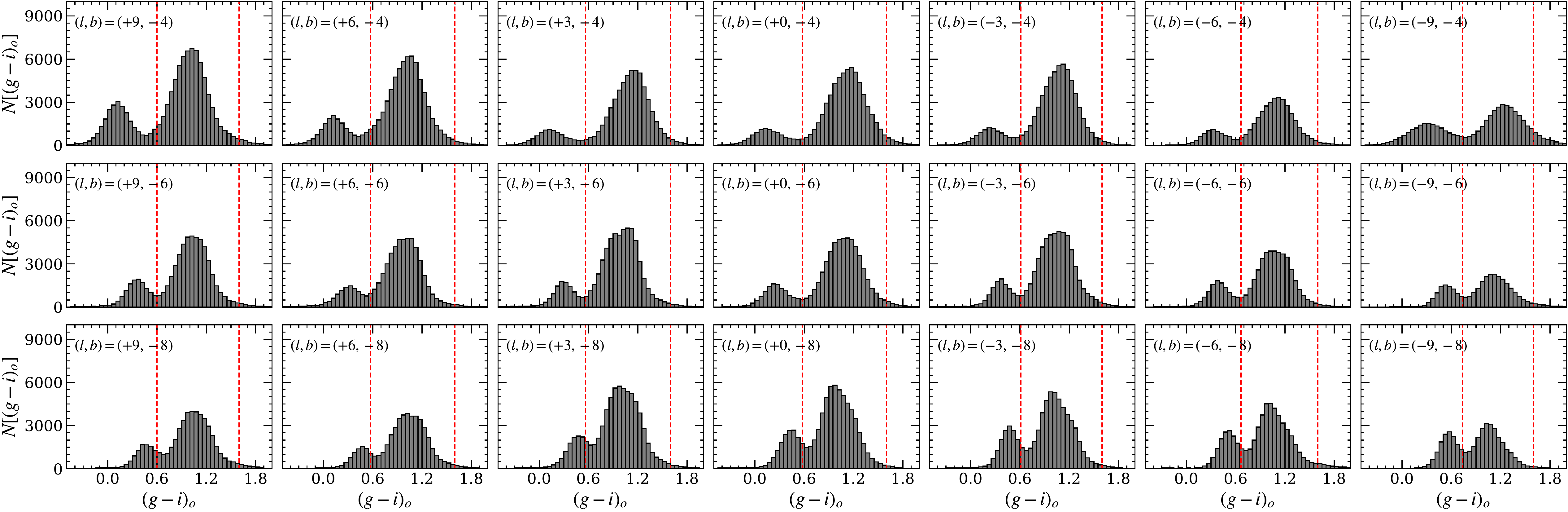}
\caption{The $(g-i)_{o}$ histograms reflect all stars in each field with 
$i_{o}$ magnitudes falling between the limits shown in 
Figure \ref{fig:app_cmd_selection_grid}.  The two populations in each panel
are the foreground main-sequence (blue group) and the bulge red clump stars
(red group).  Note that the blue main-sequence population shifts redward with
increasing distance from the plane, which simultaneously decreases the color
difference between the two groups.  However, at higher latitudes the foreground
sample is also more easily rejected by Gaia proper motions.}
\label{fig:app_cmd_selection_grid2}
\end{figure*}

Similarly, the vertical dashed red lines in 
Figure \ref{fig:app_cmd_selection_grid2} illustrate the adopted $(g-i)_{o}$ 
value used to separate the bulge red clump from foreground main-sequence stars,
based on the application of Equation \ref{eq:blue_lim}.  Recall that the 
red color limit is fixed at $(g-i)_{o}$ = 1.6 mag. for all fields.  
Figure \ref{fig:app_cmd_selection_grid2} shows that for all fields we are 
able to adequately separate the bulge red clump region from a majority of 
the foreground main-sequence stars that populate the ``blue plume" region
(e.g., see Figure \ref{fig:rc_cmd}).  Note that the relative positions of the
red clump and foreground disk populations shift in $(g-i)_{o}$ with latitude.
This is due to a combination of changing mean metallicities for the red clump
sample and the effects of applying bulge reddening values to a largely 
foreground disk population.  Note also that 
Figure \ref{fig:app_cmd_selection_grid2} includes all stars in the BDBS 
catalog, and that additional proper motion filtering further reduces the impact
of foreground contamination on our bulge red clump sample.

\bsp    
\label{lastpage}
\end{document}